\magnification=\magstep1 \overfullrule=0pt 
\advance\hoffset by -0.35truecm   
\vsize=23.1truecm
\font\tenmsb=msbm10       \font\sevenmsb=msbm7
\font\fivemsb=msbm5       \newfam\msbfam
\textfont\msbfam=\tenmsb  \scriptfont\msbfam=\sevenmsb
\scriptscriptfont\msbfam=\fivemsb
\def\Bbb#1{{\fam\msbfam\relax#1}}
\def\R{{\Bbb R}}\def\Z{{\Bbb Z}}
\def\T{{\Bbb T}}
\font\grosss=cmr8 scaled \magstep4

\def\lb{\lbrack}\def\rb{\rbrack}  \def\q#1{$\lb${\rm #1}$\rb$}
\def\bn{\bigskip\noindent} \def\mn{\medskip\smallskip\noindent}
\def\sn{\smallskip\noindent} 
\def\a{\hbox{$\cal A$}}   
\def\h{\hbox{$\cal H$}}   
\def\cedille#1{\setbox0=\hbox{#1}\ifdim\ht0=1ex \accent'30 #1%
 \else{\ooalign{\hidewidth\char'30\hidewidth\crcr\unbox0}}\fi}
\def\gaw{Gaw\cedille edzki}

\def\sbullet{\hbox{$\,{\scriptstyle \bullet}\,$}}
\def\sumlamube{\sum_{{\mathbf \lambda} ,{\mathbf \mu}}\!{}^{{}^{\beta}} }
\def\sumlamubebe{\sum_{{\mathbf \lambda} ,{\mathbf \mu}}\!{}^{{}^{\beta,b}} }
\def\sumlamubetil{\sum_{{\tilde{\mathbf \lambda}} ,
{\tilde{\mathbf \mu}}}\!{}^{{}^{\beta}} }
\def\sumlamuev{\sum_{{\mathbf \lambda} ,{\mathbf \mu}}\!{}^{{}^{\rm ev}} }
\def\sumlamuevpr{\sum_{{\mathbf \lambda'} ,{\mathbf \mu'}}\!{}^{{}^{\rm ev}} }
\def\hbn{\hfill\break\noindent}
{\nopagenumbers
\line{IHES/P/97/91 \hfill DESY 97-253}
\sn
\bn\bn\bn\bn
\centerline{
{\grosss     D-branes in Gepner models}}
\bn\bn
\vfill \centerline{by}\bn\bn \vfil
\centerline{
{\bf A.\ Recknagel}$\,{}^{*}\,$\ \  and\ \  {\bf V.\ Schomerus}$\,{}^{**}$}
\bn\bn\bn
\centerline{${}^*$Institut des Hautes \'Etudes Scientifiques,} 
\centerline{35, route de Chartres, F-91440 Bures-sur-Yvette, France}
\bn
\centerline{${}^{**}\,$II. Institut  f\"ur Theoretische Physik,
Universit\"at Hamburg,}
\centerline{Luruper Chaussee 149, D-22761 Hamburg, Germany}
\bn\bn\vfill\vfill\vfill\bn
\centerline{\bf Abstract}
\bigskip\narrower
\noindent We discuss D-branes from a conformal field theory 
point of view. In this approach, branes are described by boundary 
states providing sources for closed string modes, independently 
of classical notions.  The boundary 
states must satisfy constraints which fall into 
two classes: The first consists of gluing conditions between 
left- and right-moving Virasoro or further symmetry generators, 
whereas the second encompasses non-linear consistency conditions from 
world sheet duality, which severely restrict the allowed 
boundary states. We exploit these conditions to give explicit 
formulas for boundary states in Gepner models, thereby computing 
excitation spectra of brane configurations. From the 
boundary states, brane tensions and RR charges can also be 
read off directly. 

\bn\bn
\bn\bn
\vfill\vfill\vfill
\bigskip
\medskip 
\centerline{e-mail addresses: anderl@ihes.fr and vschomer@x4u2.desy.de}
\eject}
\pageno=1
\noindent
{\bf 1. Introduction}
\mn
D-branes were first encountered as solitonic solutions 
to the low-energy effective equations of motion of 
superstring theory, as collective excitations extending 
in $p+1$ directions of the target ("$p$-branes"), 
well-localized in the other, carrying Ramond-Ramond (RR)
charges (see \q{1} for a comprehensive review). 
Independently of these $p$-branes, Polchinski et al.\ 
\q{2} discovered that a theory of closed and open 
strings in a toroidal target can equivalently be 
described as a closed string theory with additional 
higher-dimensional objects which (via $T$-duality) 
"implement" the boundary conditions originally imposed 
on the open strings. In particular, if Neumann boundary 
conditions hold in $p+1$ directions of the target (one 
of them being time) and Dirichlet conditions in the others, 
one obtains a Dirichlet $p$-brane with $p+1$-dimensional 
world volume to which the closed strings couple. 
\hbn
Later, Polchinski showed that the Dirichlet $p$-branes 
indeed provide a "microscopic" realization of the solitonic 
$p$-branes within full-fledged string theory: They carry the 
same RR charges and their tension scales like $g_S^{-1}$ 
with the string coupling; see \q{3-5}. 
\sn
This discovery made it possible to study non-perturbative aspects of 
string theory and led to a completely new picture of both string 
theory and the quantum field theories which arise as low-energy effective 
theories from string theory. Among other applications, D-branes allow 
to test the conjecture that all the different types of string theories 
known before actually are related (by S-duality) to each other and to 
eleven-dimensional M-theory \q{6,7}. The appearance of the extra 
dimension is itself due to a  Kaluza-Klein re-interpretation of the
spectrum of D-brane bound states, which also are the basic ingredients 
of strong-weak-coupling duality tests, like of type IIB self-duality 
\q{8}. In all these considerations, the BPS property of D-branes 
is of crucial importance. 
\hbn
Many of the statements on string dualities translate into relations between 
different field theories, possibly living in different dimensions. One way 
to see this is to use the fact that the massless excitations of string-brane 
systems induce low-energy effective (gauge) field theories on the world 
volume of the D-branes;  see e.g.\ \q{8-13} and references therein. 
It was argued in \q{14} that these gauge theories at the same time 
provide an affective description of sub-string scale physics and 
geometry. 
\sn
Compared to the solitonic $p$-branes of the low-energy effective field 
theories, Polchinski's D-branes are "microscopic", conformal QFT objects. 
Nevertheless, this formulation is still phrased in terms of target 
properties; in the literature, D-branes are mainly viewed as submanifolds 
of the target space, e.g.\  as flat hyper-planes of toroidal targets 
or as supersymmetric cycles of Calabi-Yau targets \q{15-17} 
(fluctuations of the shape are induced by interactions with closed 
strings). 
\hbn
While the target point of view provides intuitive pictures and is 
very convenient for a discussion of low-energy effective physics
(including moduli spaces), one should also try to establish a more 
abstract formulation in terms of world-sheet CFTs where possible. 
One obvious reason is that string theory after all is to include 
quantum gravity and thus should not be formulated in terms of classical 
geometry only. Apart from conceptual considerations, the algebraic 
approach to string theory has its merits when it comes to producing 
concrete data in a controllable way. 
\sn
\eject\noindent
In this article, we will show how (generalized) D-branes can be 
treated, in a background independent fashion, within the CFT approach 
to string theory. We expect that the new objects cannot be entirely 
described in classical geometric terms, but just like the Dirichlet 
branes introduced by Polchinski, they correspond to non-perturbative 
sectors of closed string theories. 
\hbn
Because of the open string origin of the branes, it is not surprising 
that their CFT-formulation requires notions from boundary CFT: The 
branes bound the world sheet of the closed strings which couple to them; 
the boundary CFT then describes the (perturbative) physics of the 
system made out of closed strings and (static) branes. 
\sn
Our discussion starts from "classical" D-branes in flat toroidal 
targets in section 2, where we briefly review their CFT formulation 
in terms of coherent states associated to the Fock spaces of closed 
string modes. 
In section 3, we provide the formalism necessary to abstract from 
this simple situation and turn towards the concepts of boundary 
states and boundary CFT. 
We apply the general methods to CFTs which yield particularly 
interesting string compactifications in section 4, by constructing 
generalized D-branes for arbitrary Gepner models. 

\bn\bn
{\bf 2. Dirichlet branes as boundary states}
\mn
The CFT formulation of 
``classical'' Dirichlet $p$-branes in a flat toroidal target 
was already sketched by Polchinski  \q{3-5} and worked out in great
detail in the papers \q{18-22}. The techniques used were more or 
less known from open string theory \q{23,24}; see \q{5} 
for further references.
\sn
Let us first look at open strings propagating in a 1-torus, i.e.\ 
at the free boson $X(t,\sigma)$ taking values in a circle of 
radius $r$ and with world sheet coordinates 
$t\in\R$, $0\leq\sigma\leq\pi$.  $X(t,\sigma)$ satisfies 
the equation of motion $\triangle X(t,\sigma)=0$ where $\triangle$ 
is the two-dimensional Laplacian, but the theory
is not completely fixed before we specify some boundary
condition on the two boundaries of the open string world 
sheet $\sigma = 0, \pi$. The two obvious choices are Neumann or 
Dirichlet boundary conditions: 
$$\eqalignno{
{\rm N\ open:}\qquad\qquad\qquad &{\partial X(t,\sigma)\over \partial \sigma} =0
\qquad\qquad {\rm for}\ \ \sigma=0,\pi\qquad\qquad 
&(2.1)\cr
{\rm D\ open:}\qquad\qquad\qquad &{\partial X(t,\sigma)\over \partial t} =0
\qquad\qquad {\rm for}\ \ \sigma=0,\pi\qquad\qquad 
&(2.2)\cr}
$$
Note that, in string language, imposing Dirichlet conditions means fixing  
a constant value for the string coordinate along each boundary, 
$$
X(\sigma=0,t) = x_{\alpha}\ ,\quad\quad X(\sigma=\pi,t) = x_{\beta}\ ,
\quad\ \  {\rm for\ all}\ t\ ,
\eqno(2.3)$$
rather than prescribing arbitrary functions $x_{\alpha}(t),x_{\beta}(t)$; 
otherwise, conformal symmetry of the open string would be violated 
(see also section 3). 
\sn
The question how open string boundary conditions can be related to 
closed strings has been answered in \q{23,24}: Interacting open strings 
never come without closed strings; making the open string world sheet 
periodic in $t$ with period $2\pi\tau$, say, we obtain an open string 
one-loop diagram -- which can be viewed as a closed string tree-level 
diagram by {\sl world sheet duality}. 
\hbn
In passing to the dual world sheet, the roles of world sheet time and 
space are interchanged. The space coordinate is periodic (with period 
$2\pi$), the boundaries now are at 
$t_\alpha=0$ and $t_{\beta}=\pi/\tau$, and the conditions (2.1-3) become 
$$\eqalignno{
{\rm N\ closed:}\qquad\qquad\qquad &{\partial X(t,\sigma)\over \partial t} =0
\qquad\qquad {\rm for}\ \ t=0,\pi/\tau\ ,\qquad\qquad 
&(2.4)\cr
{\rm D\ closed:}\qquad\qquad\qquad &{\partial X(t,\sigma)\over \partial \sigma} =0
\qquad\qquad {\rm for}\ \ t=0,\pi/\tau\ ,\qquad\qquad 
&(2.5)\cr}
$$
respectively $X(t_\alpha,\sigma) = x_{\alpha}\,,\ 
X(t_\beta,\sigma) = x_{\beta}$ for all $\sigma$. 
\sn
Closed string cylinder diagrams like the one we arrived at usually 
describe a closed string propagating from a state $|v_\alpha\rangle$ at 
time $t_{\alpha}$ to a state $|v_\beta\rangle$ at time $t_{\beta}$ 
with $|v_{\alpha,\beta}\rangle$ 
taken from the Hilbert space $\h$ of perturbative closed string excitations, 
i.e.\ from the Hilbert space of the free boson CFT (defined on the plane). 
\hbn
It is therefore a natural idea to try and implement the conditions (2.4,5) 
on some ``boundary states'' $|B\rangle_N$ and $|B\rangle_D$ associated to 
$\h$ such that, e.g.\ for $t_{\alpha}=0$ 
$$
{\partial X(0,\sigma)\over \partial t}\,|B\rangle_N =0
\eqno(2.6)$$
in the Neumann case and 
$$
{\partial X(0,\sigma)\over \partial \sigma}\,|B\rangle_D =0
\eqno(2.7)$$
in the Dirichlet case. (See section 3 for a more conceptual introduction 
of the boundary states.) 
\sn
Let us try to solve (2.7) first, proceeding in analogy to \q{23,24}. 
To this end, we split the closed string coordinate into left- and 
right-movers $X(t,\sigma) = X_L(x^+)+X_R(x^-)$ with $x^{\pm}=t\pm \sigma$, 
and use the mode expansion
$$\eqalignno{
  X_L(x^+) &= {\hat x\over2} 
       + {1\over2} \bigl({\hat p\over2}+ r \hat w \bigr)\,x^+ 
+ {1\over2} \sum_{n \neq 0}
              \ {\alpha_n\over n} \, e^{-inx^+}\ , 
&(2.8)\cr
   X_R(x^-) &=   {\hat x\over2} 
       + {1\over2} \bigl({\hat p\over2} - r \hat w \bigr)\,x^- 
      + {1\over2} \sum_{n \neq 0}
              \ {\overline{\alpha}_n\over n} \, e^{-inx^-}\  . 
&(2.9)\cr}
$$
The non-vanishing commutators among the operators 
$\hat x,\,\hat p,\,\hat w,\, \alpha_n,\,\overline{\alpha}_n$ 
are 
$$    [\alpha_n,\alpha_m] \ = n\,\delta_{n,-m}  \  , \ \
      [\overline{\alpha}_n,\overline{\alpha}_m] \ = \ n\, \delta_{n,-m}\ , \ \
      [\hat x,\hat p]\ = \ i \ , $$
and we have $\alpha_n^*=\alpha_{-n}$, $\overline{\alpha}{}^*_n= 
\overline{\alpha}_{-n}$ for the 
oscillator modes, while the center of mass momentum and coordinate 
$\hat p,\,\hat x$ and the winding number operator $\hat w$  are self-adjoint. 
\hbn
The derivatives
$$ J(x^+) \ := \ \partial_{+} \, X_L (x^+) \ \ \hbox{ and }
   \quad\ 
   \overline{J}(x^-) \ := \ \partial_{-} \, X_R(x^-)
$$
generate  two commuting copies of the $U(1)$-current algebra $\a 
\cong \a_L \cong \a_R $, and we set $\alpha_0 := (\hat p/2 + r\hat w) 
\in \a_L$ and $\overline{\alpha}_0 := (\hat p/2 - r\hat w) \in \a_R$. 
The space of states of 
the free bosonic field $X$ is built up from tensor products 
$\h_g \otimes \h_{\bar g}$ of irreducible highest weight representations of 
$\a_L\otimes \a_R$, containing states $|(w,k)\rangle$ 
with the properties   
$$\eqalign{
   \alpha_0 |(w,k)\rangle  &\vphantom{\sum}= (k/(2r)+rw)\, |(w,k)\rangle 
     \equiv  g\,|(w,k)\rangle \ ,
\cr 
  \overline{\alpha}_0 |(w,k)\rangle  &\vphantom{\sum}= (k/(2r)-rw)\, 
    |(w,k)\rangle 
      \equiv  \overline{g}\,|(w,k)\rangle \ ,
\cr
     \alpha_n\, |(w,k)\rangle  &\vphantom{\sum}=  
     \overline{\alpha}_n  |(w,k)\rangle = 0 \quad 
{\rm for}\ \ n>0\ .
\cr}$$
Since $X$ takes values on a circle with radius $r$, both $\hat w$ 
and $\hat k := r \hat p$ have integer spectrum. 
\sn
With these notations we can return to our
analysis of boundary conditions. From (2.8,9), it is easy to see that 
condition (2.7) on the boundary state $|B\rangle_D$ implies that  
$$
    \hat w\, |B\rangle_D =  0 \quad \ \hbox{\rm and } \quad \ 
    ( \alpha_n -  \overline{\alpha}_{-n}) \,|B\rangle_D =  0 
\eqno(2.10)$$
for all $n\neq 0$. Using the commutation relations above, the general solution of 
these equations can be constructed as a linear combination of the following
{\sl coherent states}
$$ | (0,k) \rangle\!\rangle_D  :=  \exp \Bigl(\,\sum_{n=1}^\infty {1\over n}\,
        \alpha_{-n}\, \overline{\alpha}_{-n}\Bigr)\ |(0,k)\rangle \  . 
\eqno(2.11)$$ 
The (up to normalization) unique solution $|B(x_{\alpha})\rangle_D$ which 
satisfies the stronger requirement 
$X(0,\sigma)\, |B(x_{\alpha})\rangle_D  = x_{\alpha}\,|B(x_{\alpha})\rangle_D$ is 
$$
  |B(x_{\alpha})\rangle_D =   {1\over \sqrt{2r}}\,
 \sum_{k\in\Z} e^{-i k x_{\alpha}/r}\, |(0,k) \rangle\!\rangle_D  \ .
\eqno(2.12)$$
In the literature, (2.12) is sometimes written in the form  
$|B(x_{\alpha})\rangle_D \sim \delta(\hat x-x_{\alpha}) |(0,0) \rangle\!\rangle$, 
expressing the fact that the string coordinate takes the fixed value $x_{\alpha}$ 
along the boundary where  $|B(x_{\alpha})\rangle_D$ is placed. 
\sn
Analogously, one can implement Neumann boundary conditions (2.6) on 
a boundary state $|B\rangle_N$, see \q{23,24}. In terms of modes, (2.6) 
reads 
$$
    \hat p\, |B\rangle_N =  0 \quad \ \hbox{\rm and } \quad \ 
    ( \alpha_n +  \overline{\alpha}_{-n}) \,|B\rangle_N =  0 
\eqno(2.13)$$
for all $n\neq 0$, and the general solution is a linear combination of 
coherent states 
$$ | (w,0) \rangle\!\rangle_N  :=  \exp \Bigl(\,- \sum_{n=1}^\infty 
    {1\over n}\,
        \alpha_{-n}\, \overline{\alpha}_{-n}\Bigr)\ |(w,0)\rangle \  . 
\eqno(2.14)$$ 
Of course, Dirichlet and Neumann conditions are related by $T$-duality, 
accordingly one can regard the special solution 
$$
  |B(\tilde x_{\alpha})\rangle_N =   {\sqrt{r}}\,
 \sum_{w\in\Z} e^{-2i r w \tilde x_{\alpha}}\, | (w,0) \rangle\!\rangle_N  
\eqno(2.15)$$
as an eigenstate of the ``dual'' coordinate $\tilde X= X_L-X_R$ with eigenvalue 
$\tilde x_{\alpha}$. 
\sn
An important common property of Dirichlet and Neumann boundary states 
is that both preserve conformal invariance in the sense that 
$$
( L_n -  \overline{L}_{-n}) \,|B\rangle_D =  
( L_n -  \overline{L}_{-n}) \,|B\rangle_N = 0 
$$
where $L_n$, $\overline{L}_n$ are the left- resp.\ right-moving 
Virasoro generators of the closed string, which have the usual 
bilinear expressions in $\alpha_n, \overline{\alpha}_n$. This 
observation will be the starting point of the general considerations 
in the next section. 
\sn
It is straightforward to form boundary states describing Dirichlet 
$p$-branes in a $(d+1)$-torus out of (2.12) and (2.15): Assume that Neumann 
conditions are prescribed in the directions $X^{\mu}$, $\mu=0,\ldots,p$, 
and Dirichlet conditions in the directions $X^{\nu}$, $\nu=p+1,\ldots,d$. 
Then the tensor product 
$$ |B(p-{\rm brane})\rangle = 
|B(\tilde x^0_\alpha)\rangle_N \otimes \cdots\otimes  |B(\tilde x^p_\alpha)\rangle_N
\otimes |B(x^{p+1}_\alpha)\rangle_D \otimes \cdots\otimes |B(x^{d}_\alpha)\rangle_D 
\eqno(2.16)$$ 
describes a D $p$-brane in the $(d+1)$-torus with fixed locations $x^{\nu}_\alpha$ 
in the last $d-p$ and fixed ``dual locations'' $\tilde x^{\mu}_\alpha$ in the first 
$p+1$ directions. 
\mn
Generalization of the above construction from the bosonic case to 
boundary states describing D-branes for superstrings in a flat ``super torus'' 
is easy, too. One merely has to impose boundary conditions on the fermionic 
coordinates $\psi^{\mu}$, $\overline{\psi}{}^{\mu}$ as well, namely 
$$
\bigl( \psi^{\mu}_r \pm i \,\overline{\psi}{}^{\mu}_{-r} \bigr) 
|B\rangle_{\psi} = 0 \ ,
\eqno(2.17)$$
and to solve them in the fermionic Fock space; again, fermionic coherent 
states arise. The signs in (2.17) correspond to different 
choices of the spin structure, cf.\ \q{5,25}, the factor $i$ will 
become clear in section 3. 
\def\oh{{1\over2}}
\mn
Let us sketch how the boundary states (2.11,12) and (2.14,15) can be 
used to uncover properties of the string-brane system, first 
restricting to the one-dimensional Dirichlet case again. As mentioned 
above, the closed string cylinder diagram describes a transition 
amplitude from a boundary state $|B(x_{\alpha})\rangle_D$ at time $t_\alpha$ 
to a boundary state $|B(x_{\beta})\rangle_D$ at time $t_\beta$, with propagation 
driven by the closed string Hamiltonian $H_{\rm cl} = L_0 + \overline{L}_0 - c/12$. 
{}From the open string point of view, we calculate a one-loop diagram, i.e.\ a 
partition function $Z_{\alpha\beta}^D$ of the open string Hamiltonian 
$H_{\rm op}= L_0 - c/24$. (We neglect the usual offset $-1$ from 
normal ordering.) We have the equation
$$
Z_{\alpha\beta}^D(q) \equiv {\rm Tr}_{{\cal H}_{\rm op}} q^{H_{\rm op}} = 
{}_D\langle B(x_{\beta}) |\, \tilde q ^{\oh H_{\rm cl}} \, |B(x_{\alpha})\rangle_D
\eqno(2.18)$$
with $q = e^{2\pi i \tau}$ and $\tilde q = e^{-2\pi i /\tau}$. The right 
hand side is easily evaluated using (2.11,12) and Poisson resummation, 
see e.g.\ \q{26}:   
$$
Z_{\alpha\beta}^D(q) = {1 \over \eta(q) }\, \sum_{k\in\Z}  q^{ {1\over2}
\lb 2rk + (x_{\alpha}-x_{\beta})/\pi\rb^2 } 
\eqno(2.19)$$
with $\eta(q) = q^{1\over24} \prod_{n\geq1} (1-q^n)$. 
Similarly, for the case that Neumann conditions are prescribed at 
both ends of the open string, formulas (2.14,15) lead to 
$$
Z_{\alpha\beta}^N(q) =  {1 \over \eta(q) }\, \sum_{w\in\Z} 
 q^{{1\over2}\lb w/r + (\tilde x_{\alpha}-\tilde x_{\beta})/\pi\rb^2} 
\ \ . \eqno(2.20)$$
Neumann conditions at one end and Dirichlet at the other can of course 
also be imposed; see \q{26} for the associated partition function. 
\hbn
We observe that both partition functions are linear combinations 
of U(1) characters with positive integer coefficients. Furthermore, 
during the Poisson resummation, 
the momentum quantum number $k=rp$ in the Dirichlet coherent states (2.11)
has acquired the role of a winding number in (2.19), whereas the winding 
degree of freedom $w$ in the Neumann coherent states (2.14) has turned 
into a momentum degree of freedom in the partition function (2.20). 
\hbn
When generalized to tensor product boundary states $|B_\alpha\rangle$ 
and $B_{\beta}\rangle$ 
as in (2.16), the partition function $Z_{\alpha\beta}(q) = 
\langle B_{\beta} |\, 
\tilde q ^{\oh H^{cl}} \, |B_\alpha\rangle$ describes 
the (perturbative) excitation 
spectrum of two D $p$-branes with locations $\tilde x^{\mu}_\alpha, 
x^{\nu}_\alpha$ and  
$\tilde x^{\mu}_{\beta}, x^{\nu}_{\beta}$, interchanging (tree-level) 
closed strings 
between each other. In particular, one can read off the massless states of 
the system from $Z_{\alpha\beta}(q)$ -- and therefore from the boundary states -- 
by identifying the highest weight states of conformal dimension one. 
\hbn
We see that for large target radius $r$, the winding modes in (2.19) become 
very heavy, whereas the momentum modes in (2.20) stay light: The low-energy 
effective field theory contains fields that depend only on the Neumann directions 
of the brane, i.e.\ on its world volume coordinates. This was explained in 
\q{4,5} and further used in \q{8} to show that 
D-branes for the superstring (with Chan-Paton factors added) support 
dimensionally reduced 
gauge theories, with additional Higgs-like degrees of freedom which 
can be viewed as non-commutative transverse coordinates of the brane. 
\sn 
The CFT partition functions above -- or more generally traces with 
additional vertex operators inserted, cf.\ (3.6) below -- 
enter string amplitudes in the form 
$$
A = c \int_{{\cal M}}\! d\mu_{{\cal M}} \,Z_{\alpha\beta}(q)\ .
\eqno(2.21)$$
Here, the integral is over the moduli of the world sheet (in the simplest 
case shown, over the period $\tau$ with an appropriate invariant measure), 
and the constant $c$ contains symmetry factors of the diagram and the 
string tension, see \q{5,18-22} for more details.
\hbn 
Some data on the low-energy effective field theory can be uncovered 
without performing the integral in (2.21): As indicated above, 
the field content of the low-energy effective field theory on the 
D-brane world volume follows from the partition function 
$Z_{\alpha\beta}(q)$ alone, and  
certain couplings $\kappa_\Psi$ in this field theory  
can be  read off directly from the boundary states (up to universal prefactors 
like $c$ above). In particular, it was shown in \q{19-22} that the tension 
of the brane and its RR charges follow simply by taking projections 
$$
\kappa_\Psi  \sim \langle \Psi |\,B(p-{\rm brane})\rangle
\eqno(2.22)$$
where $|\Psi\rangle$ denotes the closed string state associated to the 
(massless) field in question; to obtain the tension, one inserts the 
graviton state, for the RR charges the corresponding anti-symmetric forms. 
Beyond that, Di Vecchia et al.\ \q{21} could even derive the solitonic 
$p$-brane solution of the low-energy effective equations of motion from 
the Dirichlet $p$-brane boundary state. 
\sn
In none of the formulas above, the closed string coupling $g_S$ shows up. However, 
when we want to compare amplitudes like (2.21) to  closed string diagrams 
describing transitions between ordinary closed string states instead of 
boundary states, we have to keep track of the $g_S$-powers  properly. 
Because of the open string origin of the boundary states, one finds \q{3,5}
that e.g.\ the D-brane tension, which can be expressed as a disc diagram, 
scales like $g_S^{-1}$ -- establishing the non-perturbative nature of 
D-branes and boundary states. 

\bn\bn
{\bf 3. Generalized D-branes, and boundary CFT}
\mn
We will now discuss methods which allow to 
abstract from the particularly simple case above and to define 
"D-branes" in CFT terms, without reference to the geometry 
of some target. (We will, nevertheless, stick to the familiar 
name for these more general non-perturbative objects of 
string theory.) The relevant 
``microscopic'' formalism to discuss such situations is
that of boundary CFT, which goes back to the work of Cardy 
\q{27}. In the first two subsections, we will set up this 
framework in a generalized form. 
\hbn
The central aim of the section is to 
explain the notion and properties of {\sl boundary states}
which contain the complete information about a (static)
D-brane. Their explicit construction is divided into two 
separate steps. First, in subsection 3.3, we obtain a linear 
space spanned by {\sl generalized coherent} or {\sl Ishibashi  
states}, which  solve linear constraints describing 
how the left-moving symmetry generators of the closed 
string  are ``glued'' to the right-moving ones on the brane. 
\hbn
Arbitrary boundary states have an expansion in a basis
of such Ishibashi states, and in the second step of the 
construction one needs to 
determine the expansion coefficients. Following 
Cardy \q{28}, we will show in subsection 3.4 that they 
are severely restricted by world sheet duality (modular 
covariance). These non-linear conditions were, so far, 
not explicitly used in the D-brane literature, but did 
play a role in previous investigations of open string 
theories \q{29}. 
\hbn
Further constraints and general features of boundary 
CFT, including the bulk-boundary operator product expansion 
(OPE) \q{30}, are sketched in subsection 3.5. We conclude the 
section with some remarks on non-rational situations.

\def\Im{{\rm Im}}\def\Re{{\rm Re}}
\bn\sn
{\bf 3.1 Boundary conformal field theory}
\mn
It is well-known that tree-level closed string theory is 
described by a CFT on the full plane (the so-called 
{\sl bulk theory}), whereas tree-level open string theory involves 
{\sl boundary CFTs} on the upper half-plane (or on 
the strip). 
In the modern context, it is the D-brane which cuts the 
string's world in half: The boundary CFT now describes the 
perturbative properties of the string-brane-system -- while, 
of course, the presence of the D-brane, or of a world sheet 
boundary, is not a perturbative phenomenon when seen from 
the original closed string. 
\sn
So let us consider a CFT in the upper half-plane ${\rm Im}\, z 
\geq 0$. Occasionally, we shall 
write $z = \exp({t} + i \sigma)$ and think of ${ t}$ as 
time and of $\sigma$ as space variable. In the interior, ${\rm Im}\, z
> 0$, the theory behaves like a usual CFT on the full plane. 
More precisely, the fields
of the boundary CFT are in 1-1 correspondence with fields of 
the associated bulk theory and locally all their structure
coincides in the sense that both theories have identical 
equal-time commutators with chiral fields and identical operator 
product expansions. In particular, there exists a stress-energy 
tensor $T_{\mu\nu}(z,\bar z)$ for $\Im \, z \geq 0$ and we 
shall require that no energy flows across the real line, i.e., 
$T_{xy}(x,0) = 0$ -- where the variables $x = \Re\, z$ and 
$y = \Im \, z$ have been used. This translates into the 
condition 
$$              T(z) \ = \ \overline T(\bar z)  \ \quad 
        \ \quad  \hbox{\rm  for } \ \ z \ = \ \bar z   
\eqno(3.1)
$$
for the standard chiral fields $T = 2(T_{xx} + i T_{xy})$ and 
$\overline T = 2(T_{xx}- iT_{xy})$. Even though they  obey the usual 
equal time commutators in the bulk including the relation  
$$
[\,T(z_1) , \overline T(\bar z_2) \,] = 0 \ \ \quad \ \hbox{\rm for } 
  \ \ {\rm Im}\, z_i\, >\, 0 \ \ \hbox{\rm and} \ \  |z_1|\, =\, |z_2|
  \ , 
$$
these two fields, defined on the upper half-plane only,  
do not suffice to construct the action of 
two commuting Virasoro algebras on the state space. However, 
due to relation  (3.1), we can still construct 
the generators $L_n^{(H)}$ of {\sl one} Virasoro algebra \q{27} by 
$$                 
   L_n^{(H)} \ := \ {{1}\over {2\pi i}} \int_C z^{n+1} T(z) dz
                  - {{1}\over {2\pi i}} \int_C \bar z^{n+1} 
                    \overline T(\bar z) d\bar z\ \  
\eqno(3.2)$$
where $C$ denotes a semi-circle in the upper half-plane
with ends on the real line. 
This implies that the
space $\h$ of states of the boundary theory decomposes
into a sum $\h = \bigoplus \h_i$ of irreducible Virasoro 
modules. 
\sn 
Primary fields $\phi(z,\bar z)$ in the boundary theory 
obey the characteristic equal time commutators 
$$ \eqalign{
[\,T(z_1), \phi(z_2,\bar z_2)\,] \ & = \ \vphantom{\sum}
\delta(z_2-z_1) \,\partial
\phi(z_1,\bar z_1) + h\,  \delta'(z_2-z_1)\, \phi(z_1,\bar z_1)
\   , \cr  
[\,\overline T(\bar z_1), \phi(z_2,\bar z_2)\,] \ & = \ 
 \delta(\bar z_2-\bar z_1)\, \overline \partial
\phi(z_1,\bar z_1) + \bar h \, \delta'(\bar z_2-\bar z_1)\, 
\phi(z_1,\bar z_1)\   , \cr}    
$$
for all $z_i$ with $\Im \, z_1\geq 0$, $\Im \, z_2 >0,$ and 
$|z_1|=|z_2|$. Note that the fields $\phi(z,\bar z)$ are 
well-defined only in the interior of the half-plane, 
which is why we will sometimes refer to them as ``bulk fields''
(not to be confused with the fields of a bulk CFT on the full 
plane). From these two relations one obtains 
$$   [\,L^{(H)}_n, \phi(z, \bar z)\,] \ = \ z^n(z \partial + 
   h (n+1)) \,\phi(z,\bar z) + \bar z^n(\bar z \overline 
   \partial + \bar h (n+1))\, \phi(z,\bar z)  \ . \eqno(3.3)
$$
Here, $h,\bar h$ are the conformal dimensions of the 
field $\phi$. We will assume below that their difference
$h-\bar h$ is (half-)integer for all primary fields in 
the theory. Using the commutation relations  ({3.3}) 
we can deduce Ward identities for correlators 
of primary fields (see, e.g., \q{27}). In turns out 
that the $N$-point functions of $N$ fields 
$\phi_{h_i,\bar h_i}(z_i,\bar z_i)$, $i=1,\ldots,N$,  
in the boundary CFT satisfy the same differential equations 
as  $2N$-point functions of fields 
$\phi_{h_i,\cdot}(z_i,\cdot)\phi_{\bar h_i,\cdot}(\bar z_i,\cdot)$, 
$i=1,\ldots,N$, with respect to the left-moving Virasoro algebra: 
Each field is accompanied by its ``mirror charge'' \q{27}.  
\sn
To conclude this subsection, let us briefly adapt the theory 
to cases in which there are other chiral fields $W(z)$ 
and $\overline W(\bar z)$ of half-integer conformal dimension 
$h_W$ besides the two fields obtained from the 
stress-energy tensor $T$. Again, the simplest 
way to guarantee the action of an extended chiral algebra 
on the state space is to assume that 
$$     W(z) \ = \ \overline W(\bar z) \ \qquad \hbox{\rm for } 
    \  z = \bar z \ .\eqno(3.4) $$ 
But since there is no equally fundamental interpretation for this 
constraint as we have for the stress-energy tensor, we 
shall relax this condition and assume the existence of a 
local automorphism $\Omega$ which acts on the space of 
chiral fields (while respecting the equal-time commutators)
such that 
$$       W(z) \ = \ \Omega (\overline W)(\bar z) \ \qquad 
    \hbox{\rm  for } \  z = \bar z \ .
    \eqno(3.5) $$
The condition  ({3.4}) is obtained when $\Omega$ is the 
trivial automorphism. In any case, we assume that $\Omega$ 
acts trivially on the Virasoro field $T$ so that the energy
flow vanishes on the real line. We can now construct modes 
$W^{(H)}_n$ 
by      
$$                 
W_n^{(H)} \ := \ {{1}\over {2\pi i}} \int_C z^{n+h_W-1} W(z) dz
             - {{1}\over {2\pi i}} \int_C \bar z^{n+h_W-1} 
                    \Omega(\overline W)(\bar z) d\bar z\ \ . 
$$
The notions of primary fields for the extended algebra and
corresponding Ward identities for correlators can be obtained
as for the Virasoro algebra above. Note that the Ward identities 
will depend explicitly on the automorphism $\Omega$. 

\bn\sn
{\bf 3.2 Boundary states}
\mn
In this subsection we demonstrate that to each boundary CFT 
one may associate a boundary state which contains all the 
information about the boundary conditions, and thus 
about (static) D-branes. Roughly speaking, the 
latter is described by an element in an appropriate extension 
of the state space $\h^{(P)}$ of the bulk theory. (In this 
subsection, we shall mark all objects of the CFT on the full plane 
by an upper index $P$ and those of the CFT on the half-plane by an 
index $H$.) 
\def\uh{^{(H)}}\def\up{^{(P)}}
\sn
To motivate the notion of boundary states, let us investigate 
the (finite temperature) correlators 
$$  \langle \phi\uh_1(z_1,\bar z_2) \cdots \phi\uh_N(z_N,\bar z_N)
   \rangle^{\beta_0} \ = \ {\rm Tr}_{{\cal H}}\bigl( e^{-\beta_0 H^{(H)}} 
    \phi\uh_1(z_1,\bar z_1) \cdots \phi\uh_N(z_N,\bar z_N)\bigr)  
\eqno(3.6)
$$
where we assume the arguments $z_i$ to be radially ordered
and where $H^{(H)} = L_0^{(H)} - c/24$.  
If the fields $\phi\uh(z,\bar z)$ 
are quasi-primary so that $\phi\uh(\lambda z,\bar \lambda \bar z) 
= \lambda^{-h} \bar \lambda^{- \bar h} \phi\uh(z,\bar z)$, the above 
correlators  ({3.6}) are (anti)-periodic in the time variable $t = 
\ln |z|$ up to a scalar factor. In this sense, the underlying 
geometry is that of a 1-loop diagram in open string theory. 
\sn
The idea now is to replace  $z,\bar z$ by 
new variables $\xi ,\bar \xi $ in terms of which the correlators 
({3.6}) may be reinterpreted as certain correlators 
for a theory on the full plane. Since the latter are necessarily
periodic in the space variable, we need to exchange the role 
of space and time in the transformation from $z$ to $\xi$. 
Let us first introduce the variable $w = \ln z = t + i \sigma$. 
Because of the periodicity properties of  ({3.6}) the theory 
essentially lives on a cylinder parameterized by $\sigma \in 
[0,\pi]$ and $t \in [t_0,t_0+\beta_0]$ with the two segments
at $t = t_0$  and $t_0+\beta_0$  identified. Now we exchange 
the role of space and time: After an appropriate rescaling by 
$2\pi/\beta_0$ we obtain a cylinder which is now periodic in 
space with period $2\pi$ and for which the time variable runs 
from $ 0$ to $2\pi^2/\beta_0$. This interchange of space and time 
is again nothing but world sheet duality. With the help of the 
exponential mapping 
$$    \xi \ =\  e^{{{2 \pi i}\over {\beta_0}} \ln z } \ \ \hbox{\rm  and} 
   \ \ \     
 \bar \xi \ =\  e^{-{{ 2 \pi i}\over{\beta_0}} \ln \bar z } \ , 
$$
the cylinder is finally mapped onto an annulus in the full 
$\xi$-plane. To rewrite the original correlators in terms of 
these new variables we make use of the transformation behaviour 
$$    \phi(\xi,\bar \xi)  =  (
{\textstyle {{d z}\over {d \xi}}})^{h}
 ({\textstyle {{d\bar z}\over {d\bar \xi}}})^{\bar h}\;  
\phi\uh(z,\bar z)\ , \quad 
T(\xi) = ({\textstyle {{d z}\over {d \xi}}})^{2}\;T\uh(z) 
+{\textstyle {c\over12}}\, \{z,\xi\}   \ \ 
\eqno(3.7)$$
for primary fields $\phi$ and the stress-energy tensor; 
$ \{z,\xi\}$ is the usual Schwartz derivative. 
It can be easily 
checked that the resulting correlators for the fields 
$\phi(\xi,\bar \xi)$ are invariant under the substitution 
$\xi \mapsto \exp (2\pi i) \xi\,$ if $\,h-\bar h$ is an 
integer (for half-integer $h-\bar h$, the fields are 
anti-periodic, i.e.\ they live on a double cover of the 
annulus). This shows (at the level of correlation functions) 
that the fields on the lhs.\ of (2.7) can be consistently defined 
on the full plane. 
\sn 
The notion of a {\sl boundary state} is introduced to 
interpret the correlators of $\phi(\xi,\bar \xi)$ 
completely within the framework of CFT on the plane. 
By definition, the boundary state of our original 
boundary CFT is a ``state'' $|\alpha\rangle$ ``in'' 
the state space $\h^{(P)}$ of the bulk theory such that 
$$ \langle \phi\up_1(\xi_1,\bar \xi_1) \cdots \phi\up_N(\xi_N, 
     \bar\xi_N) \rangle^{\beta_0} \ = \ 
    \langle \Theta \alpha |e^{- {{2\pi^2} \over {\beta_0}} \, H^{(P)}}  
     \phi\up_1(\xi_1,\bar \xi_1) \cdots 
     \phi\up_N(\xi_N, \bar\xi_N) | \alpha \rangle\ \ . 
\eqno(3.8)$$
Here, $H^{(P)} = L_0 + \overline L_0 - c/12 $ is the Hamiltonian, 
$\Theta$ the CPT-operator in the bulk theory -- and we now regard 
the fields $\phi(\xi,\bar \xi)$ as living on the full plane. 
With the help of the rule ({3.7}) and the conditions (3.5) it 
is easy to derive that $|\alpha\rangle$ obeys the following {\sl 
gluing} or {\sl Ishibashi conditions} 
$$ \bigl(\,L^{(P)}_n - \overline{L}{}^{(P)}_{-n}\,\bigr)\, 
|\alpha\rangle_\Omega = 0 
\quad\ {\rm and} \quad\ 
 \bigl(\,W^{(P)}_n - (-1)^{h_W} \Omega(\overline{W}{}^{(P)}_{-n})\, \bigr) \, 
|\alpha  \rangle_\Omega = 0 \ .
$$
Since these relations depend on $\Omega$, which we will also 
call ``gluing automorphism'' in the following, we 
mark the boundary states $|\alpha\rangle = |\alpha\rangle_\Omega$
by an extra subscript $\Omega$ throughout most of this text. The space 
of solutions to the above linear constraints is the subject of the next 
subsection.  
\sn    
For a given bulk CFT on the plane, 
one can usually find several different boundary 
theories, which give rise to 
different boundary states $|\alpha\rangle,\;|\beta\rangle$, etc. 
Looking at the expression (3.8) one is tempted to replace one
of the boundary states $|\alpha\rangle$ by some other state 
$|\beta\rangle$. Indeed, such correlators 
$$ \langle \phi\up_1(\xi_1,\bar \xi_1) \cdots \phi\up_N(\xi_N, 
     \bar\xi_N) \rangle^{\beta_0} \ = \ 
    \langle \Theta \beta |e^{- {{2\pi^2} \over {\beta_0}} \, H^{(P)}}  
     \phi\up_1(\xi_1,\bar \xi_1) \cdots 
     \phi\up_N(\xi_N, \bar\xi_N) | \alpha \rangle\ \  
\eqno(3.9)$$
make perfect sense 
and even possess an interpretation in the original boundary 
theory: When we map such correlators back into the $z$-half-plane 
they correspond to a boundary CFT for which the boundary 
condition ``jumps'' at the point $z=0$ (we will make this more 
precise in subsection 3.5). In string theory terms, this 
describes a system of two different branes exchanging 
(tree-level) closed strings. 
\hbn
Most properties of a boundary CFT
discussed so far hold true in these more general situations. 
In particular, the state space $\h_{\alpha \beta}$ of the 
boundary CFT with discontinuous boundary condition at $z=0$ 
carries an action of the chiral algebra generated by the modes 
$W_n^{(H)}$. In comparison to the usual case there is only one 
essential difference: The space $\h_{\alpha\beta}$ may fail
to contain a vacuum state $|0\rangle$ which is annihilated 
by the elements $L^{(H)}_n, n=0,\pm 1$. While this prevents
us from looking at vacuum correlators of ``bulk fields'' 
$\phi\uh(z,\bar z)$, the functions (3.6) are still well 
defined.       
 
\bn\sn
{\bf 3.3 Generalized coherent states}
\mn
We now want to solve the gluing conditions, i.e.\  find 
 ``states'' $|I\rangle\!\rangle$ on which left-
and right-moving generators of the symmetry algebra act like
$$
\bigl(\,W_n - (-1)^{h_W}\, \overline{W}_{-n}\,\bigr) 
       \,|I\rangle\!\rangle =0\ ; 
\eqno(3.10)$$
for simplicity, 
we have  dropped the superscript ${}^{(P)}$ and 
restricted ourselves to the trivial gluing 
automorphism $\Omega$ for the moment.  Assume from now on that 
the symmetry generators W$_n$ and $\overline{W}_{n}$ of the 
bulk CFT generate identical left- and right-moving chiral algebras 
$\a_L=\a_R=\a$, including the Virasoro algebra. For this situation,  
Ishibashi has shown \q{31} that to each irreducible highest 
weight representation $i$ of $\a$ on a Hilbert space $\h_i$ one 
can associate a ``state'' $|i\rangle\!\rangle$, which is unique 
up to an overall constant, such that (3.10) is satisfied. 
Using $|i,N\rangle$, $N\in\Z_+$, to denote an orthonormal basis 
of $\h_i$, one can give the (formal) expression 
$$
|i\rangle\!\rangle = \sum_{N=0}^{\infty} |i,N\rangle \otimes 
U |i,N\rangle 
\eqno(3.11)$$
for the {\sl Ishibashi state} associated to $i$; above, $U$ 
denotes an anti-unitary operator on the total chiral Hilbert 
space $\h_R=\bigoplus_i \h_i$ which satisfies the commutation 
relations 
$$
U\,\overline{W}_n = (-1)^{h_W}\,  \overline{W}_{-n}\,U
\eqno(3.12)$$
with the right-moving generators: $U$ acts like a chiral CPT 
operator. Note that $U|i,N\rangle \in \h_{i^+}$, the Hilbert 
space carrying the representation $i^+$ conjugate to $i$, 
and we will say that $|i\rangle\!\rangle$ {\sl ``couples to''} 
the Hilbert space $\h_i \otimes \h_{i^+}$. 
\hbn 
Proofs of the property (3.10) and of uniqueness (therefore of basis 
independence) can be found in \q{31,28}. Note that 
$|i\rangle\!\rangle$ is not a state in the bulk Hilbert 
space, in the same way as the coherent states describing D-branes 
of a  free boson were not ordinary states. Indeed, one 
can rewrite the coherent states, at least for Neumann 
boundary conditions, as Ishibashi states of the form (3.11), 
see \q{31}. 
\hbn
Let us remark already now that the divergences coming with 
objects like (3.11) will not cause any trouble later, 
since in all calculations to follow, $|i\rangle\!\rangle$ will 
be accompanied with ``damping operators'' like $q^{L_0-{c\over24}}$ 
with $|q|<1$. In particular, there is an {\sl inner product} between 
Ishibashi states that will be important below: 
$$
\langle\!\langle j|\, q^{L_0-{c\over24}}\, |i\rangle\!\rangle
= \delta_{i,j}\, \chi_i(q)
\eqno(3.13)$$
with $\chi_i(q)= {\rm Tr}_{{\cal H}_i} q^{L_0-{c\over24}}$ being 
the conformal character of the irreducible representation $i$; 
this follows from the expression (3.11). 
\sn
The gluing conditions (3.10) we have 
considered up to now depend only on the conformal dimensions 
of the local fields $W(z)$. In particular, we have 
$L_n = \overline{L}_{-n}$ at the boundary, as well as $\alpha_n 
= - \overline{\alpha}_{-n}$ for the modes of a  spin 1 current:
This means that the Ishibashi states (3.11) are {\sl not} suitable 
to incorporate Dirichlet boundary conditions on a compactified 
free boson. 
\hbn
To incorporate Dirichlet boundary conditions we now allow for 
non-trivial automorphisms $\Omega$ and look for 
solutions $|i\rangle\!\rangle_{\Omega}$ of 
$$
\bigl(\,W_n - (-1)^{h_W}\,  \Omega\bigl(\overline{W}_{-n}\bigr)\,\bigr) 
\,|i\rangle\!\rangle_{\Omega} \ =\ 0 \ .  
\eqno(3.14)$$ 
Such ``twisted'' Ishibashi states $|i\rangle\!\rangle_{\Omega}$ 
may be associated to any representation $i$ of $\a$. Given the 
standard Ishibashi state $|i\rangle\!\rangle \equiv |i\rangle\!
\rangle_{{\rm id}}$, the twisted one $|i\rangle\!\rangle_{\Omega}$  
is constructed as follows: 
\hbn
The outer automorphism $\Omega$ allows $\h_i$ to carry another 
inequivalent irreducible representation where elements $A \in \a$ 
act on states through $A |h\rangle := \Omega (A) |h\rangle$. The 
space $\h_i$ equipped with this new action of $\a$ is isomorphic 
(as an $\a$-module) to some representation space $\h_{\omega(i)}$ 
(this defines $\omega(i)$). We denote the isomorphism by 
$$ 
V_\Omega\,:\ \h_i \longrightarrow \h_{\omega(i)}\ \ , 
\eqno(3.15)$$
and we assume that $V_{\Omega} U = U V_{\Omega}$. Then the 
generalized coherent state 
$$
|i\rangle\!\rangle_{\Omega}:= ({\rm id} \otimes V_\Omega)\, 
|i\rangle\!\rangle
\eqno(3.16)$$
satisfies (3.14). Note that $|i\rangle\!\rangle_{\Omega}$ 
couples to the Hilbert space $\h_i \otimes \h_{\omega(i)^+}$. 
\hbn
Formula (3.13) holds for twisted Ishibashi states as well, 
provided both $|i\rangle\!\rangle_{\Omega}$ and $|j\rangle\!\rangle_{\Omega}$ 
belong to the same twisting $\Omega$. Otherwise, weighted traces of 
$V_{\Omega} V^*_{\Omega'}$ will appear, see section 4.2 for an example. 
\mn
The first important case of an automorphism $\Omega$ and its 
twisted Ishibashi states is given by the sign-flip 
$$
\Omega_{\Gamma}\,:\ J(z) \longmapsto - J(z) 
$$ 
of the U(1) current algebra, implying gluing conditions 
$ \alpha_n = \overline{\alpha}_n$ for the bulk modes at the 
boundary. The associated twisted Ishibashi states $|i\rangle
\!\rangle_{\Omega_{\Gamma}}$  are precisely the coherent states 
that implement Dirichlet conditions on a free boson on a circle. 
\hbn 
Generalization to non-abelian current algebras is in 
principle straightforward: The gluing relations (3.10) 
with $W_n=J^a_n$ can be twisted by outer automorphisms of the 
Dynkin diagram, cf.\ the detailed analysis by Kato and Okada in 
\q{32}. We would, however,  like to point out that their interpretation of 
the standard gluing conditions as ``Neumann conditions on the group 
target'' is erroneous in the non-abelian case, as can be seen by 
expressing the currents through the group valued field. The gluing 
conditions determine the (world sheet) symmetry content of the 
brane-bulk system but in general have {\sl no simple meaning} in 
terms of a $\sigma$-model target. 
\sn 
The other prominent example that will be important for us 
later on is the {\sl mirror automorphism} $\Omega_M$ of the 
$N=2$ super Virasoro algebra  
$$\eqalign{
&\lb\,L_n, L_m\,\rb\  = (n-m)\, L_{n+m} + {c \over 12} 
       (n^3-n)\,\delta_{n+m,0}\ ,
\cr
&\lb\,L_n, J_m\,\rb\ \  = -m\, J_{n+m}\ ,
\cr
&\lb\,L_n, G^{\pm}_r\,\rb\ \  = \bigl(\,{n\over2}-r\bigr)\,G^{\pm}_{n+r}\ ,
\cr
&\lb\,J_n, J_m\,\rb\ \ \  = {c\over3}\, n\, \delta_{n+m,0}\ ,
\cr
&\lb\,J_n, G^{\pm}_r\,\rb\ \  = \pm G^{\pm}_{n+r}\ ,
\cr
&\{\,G^{\pm}_r,G^{\mp}_s\,\} = 2 L_{r+s} + (r-s)\, J_{r+s} + 
      {c\over3} \bigl(r^2-{1\over4}\,\bigr)\  \delta_{r+s,0}\ .
\cr}
\eqno(3.17)$$
Above, $J_n$, $n\in\Z$, are the modes of a U(1) current, 
$G^{\pm}_r$ are spin $3/2$ superpartners of the Virasoro 
generators $L_n$ with $r \in \Z+{1\over2}$ (Neveu-Schwarz sector) or 
$r\in \Z$  (Ramond sector). This $N=2$ supersymmetric extension 
of the Virasoro algebra has many interesting properties (see \q{33} 
for an excellent reference), and it plays a central role in the CFT 
approach to superstrings. 
\hbn
The relations (3.17) admit the outer automorphism (the ``mirror map'') 
$$  
\Omega_M\,:\ J_n \longmapsto  - J_n \ ,\quad 
G^{\pm}_{r} \longmapsto  G^{\mp}_{r}\ ;
\eqno(3.18)$$
consequently, there are two possible sets of Ishibashi conditions 
\q{17}, usually called {\sl B-type conditions} for the standard gluing 
$$\eqalign{
&(L_n-\overline{L}_{-n})\,| i\rangle\!\rangle_B \;=\; \vphantom{\sum}
(J_n+\overline{J}_{-n})\,|  i\rangle\!\rangle_B= 0 \phantom{XXXXXXXX}
\cr
&(G^+_r+i\eta\,\overline{G}{}^+_{-r})\,|  i\rangle\!\rangle_B = 
(G^-_r+i\eta\,\overline{G}{}^-_{-r})\,|  i\rangle\!\rangle_B =0
\cr}
\eqno(3.19)$$
and {\sl A-type conditions} for the twisted gluing 
$$\eqalign{
&(L_n-\overline{L}_{-n})\,|i\rangle\!\rangle_A \;=\; \vphantom{\sum}
(J_n-\overline{J}_{-n})\,|i\rangle\!\rangle_A = 0 \phantom{XXXXXXXX}
\cr
&(G^+_r+i\eta\,\overline{G}{}^-_{-r})\,| i\rangle\!\rangle_A =
(G^-_r+i\eta\,\overline{G}{}^+_{-r})\,| i\rangle\!\rangle_A=0\ ;
\cr}
\eqno(3.20)$$
the sign freedom $\eta=\pm1$ is as in eq.\ (2.17). In the notation 
used above, $| i\rangle\!\rangle_B = | i\rangle\!\rangle_{{\rm id}}$ 
and $| i\rangle\!\rangle_A =| i\rangle\!\rangle_{\Omega_M}$. 
\sn
In the case of bosons with values on a two- or higher-dimensional torus 
(with equal radii), and also in some of the examples to be studied in the 
next section, tensor products of identical CFTs occur. Then there exist  
further outer automorphisms $\Omega_{\pi}$ of the chiral algebra, 
acting by permutation of identical component algebras. E.g, on a 
torus $\T^d$, one can glue the $\mu\,$th left-moving current 
$\partial X^{\mu}$ to the $\pi(\mu)\,$th right-moving current 
$\overline{\partial} X^{\pi(\mu)}$ with the help of such permutation 
automorphisms for any $\pi\in S_d$. 
Composition with the sign flip $\Omega_\Gamma$ allows to obtain permuted 
Dirichlet conditions as well. Similarly, one can model target rotations 
of the brane (or branes carrying constant electric fields) by suitable 
choices of the gluing automorphisms. 
\mn
We have obtained a complete overview of the possible Ishibashi states 
for a bulk CFT with symmetry algebra $\a\otimes\a$: To each automorphism 
$\Omega$ of $\a$ (including the identity) there is a set of gluing 
conditions (3.14) between the left- and right-moving generators; 
and to each irreducible highest weight representation $i$ of $\a$ on a 
Hilbert space $\h_i$, there is a unique (up to rescaling) Ishibashi 
state $|i\rangle\!\rangle_{\Omega}$ which ``implements'' these gluing 
conditions. This means that if we form a {\sl boundary state} 
$$
|\alpha\rangle_{\Omega} = \sum B^i_\alpha |i\rangle\!\rangle_{\Omega}
\eqno(3.21)$$
as a linear combination of (twisted) Ishibashi states, the system 
(bulk CFT + boundary state) will have $\a$ as its symmetry algebra. 
\sn
However, a bulk CFT is not specified by its symmetry algebra 
$\a_L\otimes\a_R$ alone, but it comes with a given modular 
invariant partition function on the torus; i.e., a consistent 
selection of irreducible $\a_{L,R}$-modules making up the total 
Hilbert space 
$$
\h_{\rm tot} = \bigoplus_{j,\bar\jmath} \h_j \otimes \h_{\bar\jmath}\ ,
\eqno(3.22)$$
$j\in I_L,\ \bar\jmath\in I_R$, is prescribed, too. Therefore, 
not all the (twisted) Ishibashi states $|i\rangle\!\rangle_{\Omega}$
that abstractly exist can really couple to the bulk theory: The 
condition is that the term $\h_i \otimes \h_{\omega(i)^+}$ occurs 
in (3.22). 
\sn
For the Gepner models studied below, this fact implies that 
a different number of Ishibashi states contribute to A-type 
boundary states ($\Omega=\Omega_M$ in (3.21)) than to 
B-type boundary states (($\Omega={\rm id}$ in (3.21)). 
E.g., A-type boundary states receive contributions from 
the chiral-chiral states in $\h_{\rm tot}$, whereas the 
chiral-antichiral states only contribute to B-type boundary states. 
\hbn
The coupling conditions also explain, in CFT terms, 
why there exist only Dirichlet $p$-branes with $p$ even in a 
ten-dimensional type IIA superstring theory (with 
flat toroidal target) whereas type IIB superstrings 
couple to D $p$-branes with $p$ odd. The difference between these 
two theories of closed superstrings lies in the GSO projection, which 
eliminates different left-right-combinations of fermionic states -- 
so that different Ishibashi states couple to the two bulk CFTs.
Details can be found e.g.\ in \q{25}. 

\bn\sn
{\bf 3.4 Cardy's conditions}
\mn
We will now discuss the second class of constraints on boundary 
states, the {\sl Cardy constraints}, which restrict the possible 
linear combinations (3.21) of Ishibashi states in a boundary 
state. While the Ishibashi states (to a given gluing condition) 
still form an abstract vector space, the new constraints are 
non-linear: The terminology boundary ``states'' is, therefore, even more 
misleading than for the generalized coherent states  $|i\rangle\!
\rangle_{\Omega}$, and one should rather view them as ``labels'' 
for different (non-perturbative) sectors of a bulk CFT -- much in 
the spirit of D-brane physics. 
\sn
Cardy's derivation  \q{28} of the conditions (see also \q{34}
for a good exposition) starts from a specialization of the 
setting we have discussed in subsection 3.2: 
Consider a boundary CFT with symmetry algebra $\a$ for which
the boundary condition jumps at $z=0$ from $\alpha$ to $\beta$; 
study the system at finite temperature, or make the ``time'' 
direction periodic with period $\beta_0= -2\pi i\tau$, then 
compute the partition function 
$$
Z_{\alpha\beta}(q) 
      = {\rm Tr}_{{\cal H}_{\alpha\beta}} q^{H^{(H)}_{\alpha\beta}}
\eqno(3.23)$$
where $\h_{\alpha\beta}$ is the boundary CFT Hilbert space, where  
$H^{(H)}_{\alpha\beta}= L^{(H)}_0-{c\over24}$ is the Hamiltonian in 
the $w = \ln z $ coordinate, and where $q = e^{2 \pi i \tau}$. 
This is in fact a correlator of the type (3.6), without insertions 
of bulk fields $\phi(z, \bar z)$, and as in subsection 3.2, eq.\ (3.9),  
we can re-interpret this situation and compute the same partition 
function within the bulk theory.
\hbn
This involves the interchange of space 
and time familiar from world sheet duality which,  
in terms or the parameter $\tau$,  amounts to a modular 
transformation $\tau \longmapsto -1/\tau$. Note that, in section 2, 
we have used world sheet duality merely to  ``translate'' open string 
Neumann or Dirichlet boundary conditions into conditions on closed 
string modes, following \q{23,24}. Cardy's constraints, on the 
other hand, will arise from the concrete modular 
covariance properties of the partition function.  
\sn
After the re-interpretation, the system is periodic in space 
(it lives on an annulus), and as before we assume that the 
boundary conditions are implemented by boundary states 
$|\alpha\rangle$ and $|\beta\rangle$, which sit at the ends 
of the annulus. For simplicity, let us restrict to standard 
gluing conditions (3.10). Let us furthermore assume that the 
Hilbert space of the bulk CFT on the full plane decomposes 
into irreducible representations of the symmetry 
algebra $\a\otimes\a$ as 
$$
 \h_{\rm tot} = \bigoplus_{j\in I} \h_j \otimes \h_{j^+}\ ,
\eqno(3.24)$$
corresponding to a so-called ``charge conjugate'' modular 
invariant partition function. 
\hbn
For the special case under consideration, eq.\ (3.9) yields 
$$
Z_{\alpha\beta}(q) = \langle \Theta\beta|\,
\tilde q^{{1\over2}( L^{(P)}_0+ \overline{L}{}^{(P)}_0-{c\over12})}\,
|\alpha\rangle
\eqno(3.25)$$
with $\tilde q =  e^{- 2 \pi i/ \tau}$. 
The right hand side describes free propagation from the 
boundary state $|\alpha\rangle$ at ``time'' 0 to the boundary state 
$|\Theta \beta\rangle$  at ``time'' $2\pi^2/\beta = \pi i /\tau $, 
driven by the Hamiltonian $H^{(P)}$. 
\hbn
Note that, deviating from Cardy, we have introduced the CPT operator 
$\Theta$ in the definitions (3.8,9) because the  two boundaries have
opposite orientation; the  same prescription was used in  
\q{24} in the context of open string theory. 
The boundary states of section 2 were CPT invariant, so we left out 
$\Theta$ for simplicity.
\sn
As was pointed out by Cardy, the simple identity (3.25) contains 
severe constraints on the boundary states $|\alpha\rangle$ and 
$|\beta\rangle$: We assumed that the boundary CFT on the half-plane 
has $\a$ as its symmetry algebra; therefore, the Hilbert space 
$\h_{\alpha\beta}$ decomposes into irreducible representations 
$\h_i$ of $\a$, and the partition function $Z_{\alpha\beta}(q)$ is a 
sum of characters 
$$
Z_{\alpha\beta}(q) = \sum_i n_{\alpha\beta}^i\,\chi_i(q) 
\eqno(3.26)$$
with $\chi_i(q) = {\rm Tr}_{{\cal H}_i} q^{L_0-{c\over24}}$ 
and {\sl positive integer coefficients} $n_{\alpha\beta}^i$. 
\hbn
\eject\noindent
On the other hand, we can compute the ``bulk amplitude''
with the help of the expansion (3.21) of boundary states 
into Ishibashi states and by applying the inner product 
(3.13): 
$$
\langle \Theta\beta|\,
\tilde q^{{1\over2}( L^{(P)}_0+ \overline{L}{}^{(P)}_0-{c\over12})}\,
|\alpha\rangle = \sum_i B^i_\beta B^i_\alpha \,\chi_i(\tilde q)
\eqno(3.27)$$
Implicitly, we have used  $\Theta\, B^j_\beta |j\rangle\!\rangle 
= \overline{B^j_\beta}\, |j^+\rangle\!\rangle$ -- i.e.\ we have picked  
a special normalization of $\Theta$ -- as well as the fact that 
the conjugate representation $i^+$ must occur in the index set $I$ 
of (3.24) as soon as $i$ does. 
\hbn
The modular transformation behind world sheet duality acts  
linearly on the characters, 
$$
\chi_i(\tilde q) = \sum_j S_{ij} \,\chi_j(q)
\eqno(3.28)$$
with 
$$ 
S S^* = 1\ , \quad S = S^{\,{\rm t}}\ , \quad S^2 = C
\eqno(3.29)$$
where $C_{ij} = \delta_{i,j^+}$ acts as charge conjugation. Thus, 
we obtain the final form of Cardy's constraints on the coefficients 
$B^i_\alpha$ of the Ishibashi states making up a full boundary state: 
$$
\sum_{i,j} B^j_\beta B^j_\alpha\, S_{ji} \,\chi_i(q) = 
\sum_i n_{\alpha\beta}^i\,\chi_i(q) 
\eqno(3.30)$$
for some set of positive integers $n_{\alpha\beta}^i$. Let us 
stress that multiplying an acceptable boundary state by an overall 
factor other than a positive integer in general violates (3.30). 
\sn
For a rational CFT, i.e.\ if $I$ is finite, Cardy has found 
a solution with the help of the Verlinde formula 
$$
N_{ij}^k = \sum_l { S_{il} S_{jl} S^*_{lk} \over S_{0l} }
\eqno(3.31)$$
for the fusion rules, where 0 denotes the vacuum representation. In 
Cardy's solution, the boundary states $|a\rangle$  carry the same 
labels as the irreducible representations of $\a$, i.e.\ $a\in I$, 
and their expansion into Ishibashi states is 
$$
|a\rangle =
 \sum_i {S_{ai}\over (S_{0i})^{1\over2} }\,|i\rangle\!\rangle\ .
\eqno(3.32)
$$
With (3.31) and the properties (3.29) of the modular $S$-matrix, 
it is easy to see that the partition function of the boundary 
CFT on the half-plane with boundary conditions described by 
states $|\alpha \rangle ,|\beta\rangle$ as in (3.32) is given 
by 
$$
Z_{ab}(q) = \sum_i N_{a^+ b^+}^{\,i}\;\chi_i(q)\ .
$$
Note that, because of the CPT operator which we introduced in 
(3.8,9,25), we obtain an additional conjugation on the rhs.\ 
compared to \q{28}. 
\sn
Even in this simple situation, it is not generally clear that the 
solutions (3.32) (together with integer multiples) cover all 
possible boundary states that may couple to the bulk theory. 
However, Cardy's solution provides sufficiently many independent 
states to render the coefficient matrix $B^i_\alpha$  invertible;
within the  analysis of sewing constraints (see \q{30,35}) this  
seems to be necessary for the notion of ``completeness'' proposed 
by Sagnotti et al.\ \q{36}. 

\bn\sn
{\bf 3.5 Boundary operators and bulk-boundary OPE}
\mn
We have explained above that the state space $\h^{(H)}$ admits 
the action of ``bulk fields'' $\phi(z\, \bar z)$, 
defined in the interior of the half-plane, and of 
chiral fields $W(z)$, $W(\bar z)$. It turns out, however, that 
one may introduce further {\sl boundary fields} $\Psi(x)$ which
are localized at points $x$ on the real line and are in 1-1 
correspondence with the elements of $\h^{(H)}$. 
\sn 
Let us suppose for the moment that the boundary condition does
not jump along the boundary so that $\h^{(H)}$ contains an 
$sl_2$-invariant vacuum state $|0\rangle$. Then, for any state
$|v \rangle \in \h^{(H)}$, there exists a boundary operator 
$\Psi_v(x)$ such that 
$$ \Psi_v(x) |0\rangle \ = \ e^{x L^{(H)}_{-1}} |v\rangle \ \ 
\ \quad\quad \phantom{x}
\eqno(3.33)$$  
for all real $x$. In particular, the operator $\Psi_v(0)$
creates the state $|v\rangle$ from the vacuum. Note also that 
$L^{(H)}_{-1}$ generates translations in the $x$-direction, 
i.e.\ parallel to the boundary.  If  $|v\rangle$ is a primary 
state of conformal weight $h$, it is straightforward to 
derive the commutators  
$$    [L^{(H)}_n, \Psi_v(x)] \ = \ x^n \bigl(x {{d} \over {dx}} 
      + h(n+1)\bigr) \; \Psi_v(x) \ .
$$
The existence of these extra fields $\Psi(x)$ motivates us to 
consider more general correlation functions in which the original 
bulk fields $\phi(z,\bar z)$ on the upper half-plane appear 
together with the boundary fields $\Psi(x)$. They obey a set of Ward 
identities extending those we have described in subsection 3.1 above.  
\sn
Following the standard reasoning in CFT, it is easy to conclude that 
the primary bulk fields $\phi(z,\bar z)$ give singular contributions 
to the correlation functions whenever $z$ approaches the real line. 
This can be seen from the fact that the Ward identities  describe a 
mirror pair of chiral charges placed on both sides of the boundary. 
Therefore, the leading singularities in $\phi(z,\bar z)$ are given
by primary fields which are localized at the point $x = {\rm Re}\, z$ 
of the real line, i.e.\ the boundary fields $\Psi_v(x)$. In other words,
the observed singular behaviour of bulk fields $\phi(z,\bar z)$ near
the boundary may be expressed in terms of a bulk-boundary OPE \q{30} 
$$     \phi(z,\bar z) \ \sim \ \sum_k \ (2 y)^{h_k - h - \bar h} 
        \ C^{\,\alpha}_{\phi\; k}\  \Psi_k(x)    \ \ . 
\eqno(3.34)$$
Here, $\Psi_k(x)$ are primary fields of conformal weight $h_k$
and $z= x+iy$ as before. We remark that the chiral bulk fields 
remain regular at the boundary. This was used before when we 
considered $W(z)$ and $\overline W(\bar z)$ to be defined in the 
closed upper half-plane, ${\rm Im} \, z \geq 0$. 
\sn
To guarantee locality of the boundary CFT, the coefficients in       
(3.34) must satisfy a number of sewing constraints similar to 
the familiar crossing relations in CFT. Such relations have been
worked out by Cardy and Lewellen \q{30,35} and in particular by 
Sagnotti et al.\ \q{36}. Since they involve the coefficients of the bulk
field OPEs, the sewing constraints are usually difficult to 
analyze. Nevertheless, some examples have been treated explicitly
(see \q{36}). In the investigation of boundary conditions for 
SU(2) WZW models, a particularly simple subset of sewing 
constraints was found: They involve only the coefficients 
$C^{\,\alpha}_{\phi\; 0}$ from the bulk-boundary OPE (3.34)  
-- where $0$ denotes the vacuum sector of the chiral algebra --, 
the quantum dimensions $d_i := (S_{0i}/S_{00})^{{1\over2}}$, and    
a special combination of fusing matrices and bulk OPE 
coefficients. Surprisingly, the latter turns out to be given by the 
fusion matrices so that the sewing constraints for the normalized 
coefficients $\widehat{C} {}^{\,\alpha}_{\phi_i\; 0} := d_i \, 
C^{\,\alpha}_{\phi_i\; 0}$ become
$$  \widehat{C} {}^{\,\alpha}_{\phi_i\; 0} \ 
   \widehat{C} {}^{\,\alpha}_{\phi_j\;0} \ = \  
    \sum_k \ N_{ij}^k \ \widehat{C} {}^{\,\alpha}_{\phi_k\; 0}  \ \; .
\eqno(3.35) $$
For non-diagonal SU(2) WZW models, some extra signs appear in the 
sum (see \q{36}). Unfortunately, the computation in 
\q{36} seems to make use of very particular features 
of the model. Nevertheless, one may hope that the simplicity 
of (3.35) is not accidental \q{37}. 
\hbn 
We shall show below that 
the coefficients $C^{\,\alpha}_{\phi_i\; 0}$, which are constrained 
by (3.35), are very closely related to the coefficients $B^i_\alpha$ 
in the boundary states (3.21). 
Therefore, equations of the form (3.35) could provide further 
conditions for the allowed boundary states beyond those described 
in the previous subsection. In the examples of section 4 we will, 
however, not take them into account.  
\sn
The relation between the coefficients $B^i_\alpha$ and 
$C^{\,\alpha}_{\phi_i\; 0}$ 
can be seen by analyzing the correlators (3.8) with 
only one primary bulk field $\phi_{(i^+,\omega(i))} (\xi,\bar \xi)$ 
inserted (the subscript $(i^+,\omega(i))$ refers to the transformation 
properties of $\phi(\xi,\bar \xi)$ under the action of chiral fields). 
\hbn
When we perform the limit $\beta_0 \rightarrow 0$, almost all terms 
in the boundary state $\langle \Theta \alpha|$ drop out of the 
correlation function and only the term  
$ B^0_\alpha\,\langle 0|$ proportional to the 
vacuum state survives. Thus, the dominant contribution to the 
1-point function for small $\beta_0$ is given by 
the correlator 
$$
\langle 0|\,\phi_{(i^+,\omega(i))} (\xi,\bar \xi) 
\,|\alpha\rangle  
= {B^i_\alpha \over (\xi\bar\xi - 1)^{h_{i^+}+h_{\omega(i)}}}
$$ 
where the rhs follows from 
invariance of both the vacuum and the boundary state under 
conformal transformations generated by $L_n - \overline{L}_{-n}\,$.
\hbn
To relate this term to the bulk-boundary OPE, we compute the same 
correlator in the $z$-plane and focus on the leading contribution 
when $z$ approaches the real line -- i.e.\ when $\xi(z)$ approaches 
the unit circle. 
We can insert the bulk-boundary OPE (3.34) to evaluate the correlator 
(3.6) in this particular regime. This leaves us with a sum 
of traces of boundary fields $\Psi_i$ multiplied by the 
coefficients of the bulk-boundary OPE, some factors 
depending on $y = {\rm Im}\, z$ as well as Jacobians from 
the coordinate transformation as in (3.7).  
When we let $y \rightarrow 0$, 
the term containing the trace of the identity field 
and the corresponding factor $C^{\,\alpha}_{(i^+,\omega(i))\;0}$ 
dominates (assuming that the CFT is unitary), with the same 
dependence on $\xi(z)$ as on the plane. 
The trace yields a factor $(B^0_\alpha)^2$, and 
comparison of the two different computations gives 
$$      C^{\,\alpha}_{(i^+,\omega(i))\; 0} \ = \ 
        {{B^{i}_\alpha} \over {B^0_\alpha}} \ \ . 
\eqno(3.36)$$ 
This shows that the coefficients in front of the identity field
in relation (3.34) determine the boundary state 
$|\alpha \rangle_\Omega$. On the other hand, using the sewing 
relations derived in \q{35,36}, one can also reconstruct 
the remaining coefficients $C^{\,\alpha}_{(i,\omega(i)^+)\; k}$ from the 
knowledge of the boundary state $|\alpha \rangle_\Omega$. 
\sn
Relation (3.36) allows us to make contact with an apparently 
different definition of boundary states which was used  in 
\q{30}: Instead of mapping the upper half-plane (with coordinate $z$) 
to an annulus in the $\xi$-plane as above, we can map it to the 
complement of the unit disc in the $\zeta$-plane via 
$$
\zeta = {1-iz \over 1+i z}\quad\quad {\rm and}  \quad\quad
\bar\zeta = {1+i\bar z \over 1-i \bar z}\ ;
$$
given a CFT on the upper half-plane with boundary condition $\alpha$, 
the boundary state for the CFT on the $\zeta$-plane is defined
in terms of zero-temperature correlators 
$$
 \langle \phi\uh_1(z_1,\bar z_2) \cdots \phi\uh_N(z_N,\bar z_N)
   \rangle_{\alpha} \ = \ \langle 0 |\, \phi\up_1(\zeta_1,\bar \zeta_1) 
\cdots \phi\up_N(\zeta_N,\bar \zeta_N)\,|\alpha\rangle\ ; 
\eqno(3.37)$$
note that the vacuum state is inserted at $\zeta=\infty$ and that, 
again, the transformation law (3.7) is to be applied. 
It is easy to verify that the Ishibashi conditions arising from 
(3.37) coincide with the previous ones, and relation (3.36) 
can be derived from the bulk boundary OPE and the large $|\zeta|$ 
behaviour of (3.37), see \q{30}. Therefore, (3.37) and the 
procedure of section 3.2 lead to the same boundary states. 
\sn
While (3.37) is certainly more convenient when dealing 
with sewing constraints, our previous definition contains 
Cardy's conditions as a special case. Moreover, it is not 
obvious how to generalize the zero temperature definition 
to
the case where the 
boundary condition jumps at the origin $z=0$ from $\beta$ to 
$\alpha$. We therefore return to the setting of section 3.2
and conclude this section with some brief remarks on discontinuous 
boundary conditions. 
\sn
As we have explained before, the state space 
$\h_{\alpha \beta}$  of such a boundary CFT does not contain a 
vacuum state so that one cannot construct vacuum correlators. 
Nevertheless, $\h_{\alpha\beta}$ does contain various primary 
states (with respect to the action of the modes $W^{(H)}_n$)
which can be inserted as incoming and outgoing states 
into the correlation functions, instead of the vacuum. 
These correlation functions may 
be interpreted as correlators of bulk fields with two extra
boundary fields $\Psi^{\alpha\beta}$ inserted at $x=0$ and
$x=\infty$. In this sense one says that non-vanishing 
correlators of a boundary theory with jumps in the boundary
condition necessarily contain some boundary fields 
$\Psi^{\alpha \beta} (x)$ (``boundary condition changing 
operators''). Operator product expansions for such more 
general boundary fields along with the associated 
sewing constraints have been described in \q{35,36}. 
\mn
Boundary CFTs have a very rich structure which in 
various respects seems to reach beyond that of their 
plane counterparts. There are many general questions 
which would be interesting to study. As for D-brane 
physics, this might lead to a deeper understanding of 
the non-perturbative string sectors. E.g., it is very 
tempting to interpret the bulk-boundary OPE (3.34) 
as kind of an explicit $S$-duality transformation   
expressing closed string modes -- i.e.\ excitations 
of the fundamental degrees of freedom in string 
theory -- through fields on the brane -- i.e.\ through 
excitations of solitonic degrees of freedom. 

\bn\sn
\eject\noindent
{\bf 3.6 Remarks on the non-rational case}
\mn
The CFTs relevant for string theory are always non-rational when 
regarded with respect to  the (super) Virasoro algebra alone, 
simply because the critical dimension is 10 or 26. For 
non-rational (boundary) CFTs, conditions (3.30) for acceptable 
boundary states continue to hold, now with infinite summations over
the character indices $i$ and $j$, but it is usually very difficult 
to obtain general solutions because the modular transformation properties 
of non-rational characters are poorly understood. 
\hbn
If we want to use the constraints from world sheet duality to 
determine boundary states, we can often exploit the presence of  bigger 
symmetry algebras. E.g., while the U(1)${}^{26}$ current algebra of 
bosonic string theory on a 26-torus is not big enough to make the CFT 
rational, its characters are sufficiently well under control so as to 
render Cardy's conditions tractable: In section 2, we have, without 
mentioning them, solved eqs.\ (3.30) by taking the correct linear 
combinations (2.12) of the Dirichlet coherent states (2.11) so as 
to obtain integer coefficients in the partition function (2.19). 
The exponential prefactors in front of the coherent states 
had an alternative interpretation as a  $\delta$-function 
$\delta(\hat x^\mu-x^\mu)$ inserted into the boundary state 
in order to fix the location of the D-brane in $\mu$-direction. 
\sn
We conclude that, as is common in string theory, geometrical properties in 
the target follow from world sheet conditions. Moreover, {\sl D-brane moduli},  
like the brane's location, can arise as free parameters in the solutions to 
those constraints. This situation can e.g.\ occur if there are different 
representations with the same character -- as is usually the case 
in a non-rational theory (often with infinitely many such ``degenerate'' 
representations). Let us recall in passing that other geometric moduli, 
like the orientation of the brane in the target, enter the CFT description 
in a different way, namely as parameters (continuous or discrete ones) 
of the gluing automorphisms $\Omega$ giving rise to families of 
Ishibashi states -- cf.\ the remarks in section 3.3.  
\sn
In other examples of string compactifications like the Gepner models, 
much bigger symmetry algebras are realized on the perturbative 
string spectrum, and with respect to these algebras the CFTs in 
question become indeed rational. 
Below, we will be content with constructing ``rational boundary 
states'' for Gepner models, i.e.\ boundary states  which preserve the 
bigger symmetry. But since, from the superstring point of view,  
only the super conformal invariance (and not an extended symmetry) 
is indispensable, we have to try and justify our procedure:
\hbn
First of all, the rational boundary states are also 
boundary states for the smaller super Virasoro algebra: Ishibashi
conditions for a big symmetry algebra $\a$ imply those for any subalgebra, 
and upon decomposing the characters of irreducible $\a$-representations into 
(possibly infinite) sums of characters of super Virasoro representations it 
is easy to see that (3.30) is satisfied also in terms of the ``smaller''
characters. 
\hbn
Second, there are indications that branes found by geometric methods 
typically preserve more symmetries ``than necessary''. This is already 
true for the classical D-branes on a flat torus, which respect the U(1)s 
and not just the diagonal Virasoro algebra; we will encounter 
the same phenomenon in the simplified Gepner model example studied in 
section 4.2 below.
\hbn
Finally, in the special situation of $N=2$ superconformal models, 
which is most important for superstrings, certain structures are 
independent of whether they are looked at from an extended symmetry 
or from a mere super Virasoro point of view. The prominent example is 
the chiral ring \q{33}, which can e.g.\ be ``transported'' through the 
whole moduli space of marginal deformations without its constituents ever 
changing. We may expect that, in a similar fashion, at least some 
properties of our boundary states are independent of rationality. 
This view is supported by the arguments given by Ooguri, Oz and Yin 
in \q{17}.
\sn
Nevertheless, constructing boundary states in a truly non-rational 
setting remains a challenge -- not only because of the D-brane 
moduli mentioned above, but also in order to clarify the status 
of non-geometric ``D-branes'' (i.e.\ to prove or disprove their 
existence). 
\mn
Before we conclude this section, we would like to stress the
following point:  Performing an explicit test of Cardy's conditions 
for a set of tentative boundary states $|\alpha\rangle$ is {\sl not} just 
a tedious exercise whose only outcome is to dismiss $|\alpha\rangle$ 
or not. The quantity $Z_{\alpha\beta}(q)$ computed in this process 
contains much of the physics of the D-brane configuration $(\alpha\,\beta)$, 
as it describes its whole (perturbative) excitation spectrum. 
In particular, the massless field content (in the string sense) of 
the low-energy effective field theory associated to the brane 
configuration is completely explicit. 
{}From the boundary state itself, one can immediately read off 
the couplings to massless closed string modes; thus, coupling 
constants of the low-energy effective field theory like 
brane tension and RR charges follow by simply ``sandwiching'' 
the boundary state with the corresponding closed string states. 
This was already recalled for the case of classical branes in 
section 2, but equally holds for arbitrary boundary states. In 
particular, ``generalized tensions'', too,  behave like 
$g_S^{-1}$, due to their open string tree-level origin. 

\bn\bn
{\bf 4. Boundary states for Gepner models}
\mn
In this section, we apply the general methods outlined above 
to a class of conformal field theories which are of particular 
interest for string theory: The Gepner models \q{38} provide an 
algebraic formulation of supersymmetric (or heterotic) string 
compactifications {}from 10 to lower dimensions, formulated in 
terms of rational ``internal'' CFTs. Since, for a long time, Gepner 
models were considered to be the most relevant string vacua as far 
as phenomenology is concerned, it is of some interest to search 
for D-branes in these models. In particular, we expect these 
non-perturbative string sectors to have consequences for dualities in 
gauge theories, too.  
\sn
We first need to review Gepner's construction of superstring vacua 
from superconformal minimal models, were it only to set up notation. 
In subsection 4.2, we will briefly discuss boundary states in 
a simplified situation, where we can in particular compare our 
construction to geometry-inspired results of Ooguri, Oz and 
Yin \q{17}. After that, in subsection 4.3, we present formulas for 
A-type and B-type boundary states in arbitrary Gepner models.

\bn\sn
{\bf 4.1 Gepner models}
\mn
In \q{38}, Gepner introduced a construction of supersymmetric 
string compactifications based on the minimal models 
of the $N=2$ super Virasoro algebra. In this algebraic approach, 
the compactification is not achieved by ``curling up'' $10-D$ 
dimensions of ten-dimensional flat space-time into a compact 
Calabi-Yau manifold, but rather by replacing the $10-D$ free superfields 
with some ``internal CFT'' of central charge $15-3D/2$ such that certain 
conditions are met, most of which serve to maintain space-time 
supersymmetry (\q{38}; see also \q{39} for a useful discussion):
\mn
\item{(1)} The internal CFT must at least have $N=2$ world sheet 
supersymmetry.
\item{(2)} The total U(1) charges must be odd integers for both left 
and right movers -- here, total refers to internal charges plus 
charges of the $D-2$ free external superfields associated to 
transversal uncompactified directions; this condition implements 
the (generalized) GSO projection. 
\item{(3)} The left-moving 
states must be taken from the NS sectors of each subtheory (external 
and, in Gepner's models, various internal sub-theories) or from 
the R sectors in each subtheory; analogously for the right-moving 
states. 
\item{(4)} The torus partition function must be modular 
invariant. 
\mn
To build concrete string compactifications, Gepner used tensor products 
of $N=2$ minimal 
models with levels $k_j$, $j=1,\ldots,r$, whose central charges sum up 
to the desired value of the internal CFT,
$$c_{\rm int} 
= 12 -{3\over2} d\ ; 
\eqno(4.1)$$ 
12 appears because of the light cone gauge, and we assume for 
later convenience that $d=D-2$ is equal to 2 or to 6. 
\sn
The (anti-)commutation relations of the  $N=2$ super Virasoro algebra
were given in (3.17). Its minimal models have central charges 
$$
    c \,= \, {3k \over k+2}\ ,\quad k=1,2,\ldots\ .
\eqno(4.2)
$$
with $k=1,2,\ldots\,$, each possessing only a finite number of unitary 
irreducible highest weight representations. For carrying out the 
GSO projection onto states of definite fermion number, one needs 
to consider representations of the bosonic subalgebra; these 
are labeled by three integers $(l,m,s)$ with
$$ 
l=0,1,\ldots,k\ ,\quad 
m= -k-1, -k, \ldots,k+2\ ,\quad
s=-1,0,1,2 
\eqno(4.3{\rm a})$$ 
and 
$$ 
   l + m + s \ {\rm even} \ .
\eqno(4.3{\rm b})$$
Triples $(l,m,s)$ and $(k-l,m+k+2,s+2)$ give rise to the 
same representation (``field identification''). 
\sn
The conformal dimension $h$ and charge $q$ of the highest 
weight state with labels $(l,m,s)$ are given by  
$$\eqalignno{
h^l_{m,s} &=   { l(l+2) - m^2 \over 4(k+2)} + {s^2\over 8}\ ({\rm mod}\,1)\ ,
&(4.4)\cr
q^l_{m,s} &=  { m \over k+2} - {s\over 2}\ ({\rm mod}\,2)\ ;
&(4.5)\cr}$$
for many purposes, it is sufficient to know $h$ (and $q$) up to (even) 
integers. The exact dimensions and charges of the highest weight 
state in the representation $(l,m,s)$ can be read off (4.4,5) if one 
first uses the field identification and the transformations 
$(l,m,s)\longmapsto (l,m+k+2,s)$ and  $(l,m,s)\longmapsto (l,m,s+2)$ to 
bring $(l,m,s)$ into the  
{\sl standard ranges} 
$$
l = 0,1,\ldots,k\ ,\quad |m-s| \leq l\ ,\quad s = -1,0,1,2\ ,
\quad l+m+s\ \  {\rm even}
\eqno(4.6{\rm a})$$
or 
$$
l= 1,\ldots,k\ ,\quad m = -l\ , \quad s = -2 \ .\phantom{xxxxxxxxxx}\ 
\eqno(4.6{\rm b})
$$
see \q{40,41}. 
Representations with an even value of $s$ are part of the NS-sector, 
while those with $s=\pm 1$ belong to the R-sector. Within
the two sectors, representations can be grouped into
pairs $(l,m,s) \& (l,m,s+2)$ which make up a full $N=2$ super 
Virasoro module, with all states in a bosonic sub-representation 
having the same fermion number modulo two. 
\mn
In order to write down partition functions that satisfy all 
the requirements listed above and therefore describe 
superstring compactifications, Gepner formed tensor products 
of $N=2$ minimal models such that the central charges add up 
to a multiple of three, see (4.1,2), then adjoined external fermions 
and bosons, and finally employed an orbifold-like procedure which 
enforces space-time supersymmetry and modular invariance \q{38,39,40}.  
\def\mathbf{}
\sn
We need some further notation before we can state Gepner's result. 
For a compactification involving $r$ minimal models, we use 
$$
{\mathbf \lambda} := (l_1,\ldots,l_r)\quad {\rm and}\quad 
{\mathbf \mu}:= (s_0; m_1,\ldots,m_r;s_1,\ldots,s_r)
\eqno(4.7)$$ 
to label the tensor product of representations: $l_j,\,m_j,\,s_j$ 
are taken from the range (4.3), and 
$s_0=0,2,\pm1$ characterizes the irreducible representations of the 
SO$(d)_1$ current algebra that is generated by the $d$ external 
fermions (the latter also contain an $N=2$ algebra for each even $d$ and, 
again, the NS-sector has $s_0$ even). \hbn
Accordingly, we write 
$$
\chi^{{\mathbf \lambda}}_{{\mathbf \mu}} (q) \;
:= \chi_{s_0}(q) \chi^{l_1}_{m_1,s_1}(q) \cdots 
 \chi^{l_r}_{m_r,s_r}(q)
\eqno(4.8)$$
with $\chi^{l}_{m,s}(q) = {\rm Tr}_{{\cal H}^l_{m,s}} q^{L_0-{c\over24}}$ 
etc.\ for the conformal characters of these tensor products
of internal minimal model and external fermion representations; 
we refer to \q{38} and \q{42} for the explicit expressions.  
\sn
We introduce the special $(2r+1)$-dimensional vectors $\beta_0$ with all 
entries equal to 1, and $\beta_j$, $j=1,\ldots,r$, having zeroes 
everywhere  except for the 1st and the $(r+1+j)$th 
entry which are equal to 2. 
\hbn
Consider the following products
%
of  $2\beta_0$ and  $\beta_i$ with a vector ${\mathbf \mu}$ 
as above: 
$$\eqalignno{ 
2\beta_0 \sbullet {\mathbf \mu} &:= 
-{d\over2} {s_0\over2} - \sum_{j=1}^r {s_j\over2} 
+ \sum_{j=1}^r {m_j\over k_j+2}\ ,
&(4.9)\cr
\beta_j \sbullet {\mathbf \mu} &:= -{d\over2} {s_0\over2} -{s_j\over2} \ .
&(4.10)\cr}$$
It is easy to see that $q_{\rm tot} := 2\beta_0 \sbullet {\mathbf \mu}$
is just the total U(1) charge of the highest weight state in 
$\chi^{{\mathbf \lambda}}_{{\mathbf \mu}} (q)$, so that the projection 
onto states with odd $2\beta_0 \sbullet {\mathbf \mu}$ will implement 
the GSO projection. Similarly, restricting to states with 
$\beta_i \sbullet {\mathbf \mu}$ integer ensures that only states in 
the tensor product of $r+1$ NS-sectors (or of $r+1$ R-sectors) are 
admitted (recall that we assumed $d=2$ or $d=6$). 
\sn
Modular invariance of the partition function can be achieved if 
the above projections are accompanied by adding ``twisted'' sectors
(in a way similar to orbifold constructions). To state Gepner's 
result, we put $K := {\rm lcm}(4,2k_j+4)$ and let 
$b_0 \in \{0,1,\ldots,K-1\}$, $b_j\in\{0,1\}$ for $j=1,\ldots,r$. 
Then the partition function of a Gepner model describing a 
superstring compactification to $d+2$ dimensions is given by 
$$
Z^{(r)}_G (\tau,\bar \tau) = {1\over2} 
{({\rm Im}\,\tau)^{-{d\over2}}\over |\eta(q)|^{2d}}
\sum_{b_0,b_j} \sumlamube 
(-1)^{s_0} \,\chi^{{\mathbf \lambda}}_{{\mathbf \mu}} (q)
\,\chi^{{\mathbf \lambda}}_{{\mathbf \mu}+b_0\beta_0+b_1\beta_1+\ldots +b_r\beta_r} 
(\bar q)
\eqno(4.11)$$
where $\sum \!{}^{{}^{\beta}}$ means that we sum only over those 
${\mathbf \lambda}, {\mathbf \mu}$ 
in the range (4.3) which satisfy $2\beta_0 \sbullet {\mathbf \mu}\in2
{\Z}+1$ and $\beta_j \sbullet {\mathbf \mu} \in {\Z}$.
The summations over $b_0, b_j$ introduce the twisted sectors corresponding 
to the $\beta$-restrictions so that, in particular, the Gepner partition 
function is non-diagonal. The sign is the usual one occurring in (space-time) 
fermion one-loop diagrams. The $\tau$-dependent factor in front of the sum 
accounts for the free bosons associated to the $d$ transversal dimensions of 
flat external space-time, while the $1/2$ is simply due to the field 
idenfication mentioned after (4.3). Using the modular transformation 
properties of the SO$(d)_1$ and minimal model characters, whose S-matrices 
are 
$$\eqalignno{
S^{\,{\rm f}}_{s_0,s'_0} &= {1\over2}\; e^{- i\pi {d\over2} {s_0s'_0\over2}}\ ,
&(4.12)\cr 
S^{\,k}_{(l,m,s),(l',m',s')} &= {1\over\sqrt{2} (k+2)} \;  
\sin \pi {(l+1)(l'+1)\over k+2} \,
e^{i\pi {mm'\over k+2}} \,e^{-i\pi {ss'\over2}}\ ,
&(4.13)\cr}$$
Gepner could prove that (4.11) is indeed modular invariant. 
\mn
Although we will work only with the above $(2,2)$-superstring compactifications, 
let us mention that originally one of the most interesting features of Gepner's 
construction was that it is straightforward to convert (4.11) into a 
(modular invariant) partition function describing a heterotic string 
compactification, see \q{38}. One simply has to replace the right-moving 
SO$(d)_1$ characters by those of 
SO$(d+8) \times $E$_8$ (with a suitable permutation). The massless spectrum 
of heterotic Gepner models then contains fermions which transform in the 
27 (and $\overline{27}$) representations of E${}_8$ and which may be 
interpreted as \hbox{(anti-)}generations of a supersymmetric GUT model derived 
as the low-energy limit of the string compactification, 
see e.g.\ \q{38,43,40}. 
\sn
In a somewhat suprising development, it turned out that Gepner's purely 
algebraic construction is intimately related to geometric string 
compactifications on certain (complete intersection) Calabi-Yau 
manifolds. E.g., the number of generations and anti-generations
computed from Gepner's partition function agrees with (or can at least 
be related to) those found in CICY-compactifictions, where they are 
given by the dimensions of certain Dolbault cohomologies. The connection 
was made more precise by the work of Greene, Witten and other authors, see 
\q{39} for a useful review. 
\hbn
An important subject that arose from the connection between 
compactifications on Calabi-Yau manifolds and the CFT 
vacua constructed by Gepner is mirror symmetry. Simply using the  
mirror automorphism (3.18) of $N=2$ superconformal field theory
in the right-moving sector, 
one can revert the role of (chiral,chiral) and (chiral,anti-chiral) 
primaries; moving to the Calabi-Yau manifolds, however, this corresponds 
to a completely non-trivial map between topologically distinct 
manifolds. This observation, made in \q{44,45}, see also \q{33} 
for a precursor, has had important consequences for algebraic geometry 
as well as for the interpretation of string theory, see e.g.\ \q{46,39} 
for details and further references. 

\bn\sn
{\bf 4.2 A simplified example}
\mn
We will now turn towards the construction of boundary states for the 
superstring compactifications reviewed above, but before we treat 
Gepner models in full generality, let us discuss the simplest case
in a simplified variant: We want to study the tensor product of 
three $k=1$ minimal models, each restricted to the NS sector, then 
orbifolded to integer U(1) charge. In contrast to the Gepner models, 
where $q_{\rm tot}$ must be odd, we do not perform the GSO-projection 
and thus still deal with a true CFT with the vacuum sector in. 
\hbn
The reasons for starting with this simple case are two-fold: 
First, notations are much lighter than for general Gepner
models. Second, and more importantly, this model can be written 
as a $\sigma$-model compactification on a 2-torus with $\Z_3$-symmetry;  
thus it has  a free field description so that boundary states have
a direct geometric interpretation, which was used in the 
treatment in \q{17}. Our construction does not rely on 
any geometric input, and in particular 
does not use the coherent Ishibashi states of free bosons 
or fermions, instead we will require our boundary states 
to render the resulting boundary CFT rational. We will 
see that the free complex fermion re-emerges naturally 
in the spectrum of the open string partition function 
determined by these ``rational boundary states'', 
which suggests that they are the right objects to start 
with. 
\mn
The first minimal model in the series (4.2) has $c=1$ and can be 
regarded as a free boson compactified on a circle with a special radius.  
The irreducible representations of the 
(here: full) $N=2$ superconformal algebra are specified by the conformal 
dimension $h$ and the charge $q$ 
of the highest weight vectors, in the NS sector given by 
$$
|(h,q)\rangle\ = \ \   |(0,0)\rangle\ ,\quad |(1/6,1/3)\rangle\ ,\quad
|(1/6,-1/3)\rangle\  .
$$
We will in the following simply use the charge 
enumerators $a=0,1,2$ (taken modulo 3) to label these representations. 
Note that $|0\rangle$ and $|1\rangle$ are chiral, $|0\rangle$ 
and $|2\rangle$ anti-chiral states. 
\hbn
The $N=2$ characters in the NS sector close under the modular 
transformation $S\,:\ \tau \longmapsto -1/\tau$, and the elements 
$S_{ab}$ in $\chi_a(\tilde q) = \sum_b  S_{ab} \chi_b(q)$ 
are given by 
$$
S_{ab} = {1\over \sqrt{3}}\; \omega^{-ab}
\eqno(4.14)$$ 
with $\omega = e^{2\pi i/3}$ and $a,b = 0,1,2$ -- as follows from 
(4.13), or more directly by diagonalizing the (simple current) fusion rules 
of the $k=1$ model. 
\sn
We take the tensor product of three $k=1$ minimal models 
(each with diagonal partition function, here restricted to 
the NS sector) and orbifold this CFT with respect to the group 
generated by 
$$
g=  \exp\{2 \pi i (J_0^{\rm tot} + \overline{J}_0\!{}^{\rm tot})\}
$$
where $J_0 = J_0^{(1)}+  J_0^{(2)}+ J_0^{(3)}$ is the total left-moving  
charge of the $(k=1)^3$ theory. The resulting partition function is 
\q{38,47,45}
$$
Z(q,\bar q) = \sum_{x=0,1,2} 
\sum_{{{\bf a}\atop {\scriptscriptstyle \Sigma a_i \equiv 0 ({\rm mod}\,3)}}} 
\!\!\chi_{{\bf a}}(q) \,
\chi_{{\bf a}+(x,x,x)}(\bar q)\ ;
\eqno(4.15)$$
the (non-diagonal) terms with $x\neq 0$ are the ``twisted sectors'', 
and the restriction on the ${\bf a} = (a_1,a_2,a_3)$ summation ensures 
that only states with integer charge occur in the orbifolded theory. 
Note that, since  $h_L-h_R$ may be half-integer,  (4.15) is only invariant 
under the subgroup of the modular group generated by $S$ and $T^2$ 
(this is already the case for the diagonal $k=1$ partition functions 
we started from). 
\sn
The maximal chiral symmetry algebra of the orbifold theory 
contains the diagonal (``total'') $N=2$ super 
Virasoro algebra, but also the bigger algebra $\a^{\otimes}$ generated 
by the three super Virasoro algebras of the individual tensor factors 
(the $g$-action commutes with all these generators). While the simple 
$c=3$ orbifold theory is rational with respect to  $\a^{\otimes}$, it is 
already non-rational with respect to the diagonal super Virasoro algebra.
\mn
If we want to construct boundary states $|\alpha\rangle$ 
of the orbifolded $(k=1)^3$ model, we can require that not only (one of) 
the diagonal $N=2$ super Virasoro algebra(s) is preserved -- in the 
form of A-type or B-type boundary conditions, see eqs.\ (3.20,19) --, 
but that the system (bulk CFT + boundary state)  in fact enjoys the full 
$\a^{\otimes}$-symmetry. This makes  Cardy's world sheet 
duality conditions tractable -- without loosing too much interesting 
structure, as comparison with the free field approach will show. 

\mn
We start with boundary states satisfying A-type conditions (3.20)
in each subtheory. Out of the 27 sectors of the orbifold theory (4.15) 
only the 9 untwisted ones ($x=0$) can contribute because of the 
charge constraints $q^{(i)} =  \overline{q}{}^{(i)}$ for $i=1,2,3$. 
\hbn
Thus, we make the following ansatz for an A-type 
$\a^{\otimes}$-preserving boundary state of the orbifolded 
$(k=1)^3$ model in the NS sector: 
$$
|\alpha\rangle_A = {1\over 3^{1/4}} 
\sum_{{{\bf a}\atop {\scriptscriptstyle \Sigma a_i \equiv 0\; ({\rm mod}\,3)}}} 
 \omega^{-a_1\alpha_1-a_2\alpha_2-a_3\alpha_3}
|{\bf a} \rangle\!\rangle_A 
\eqno(4.16)$$                                
where $|\alpha\rangle = |(\alpha_1,\alpha_2,\alpha_3)\rangle $ is labeled 
by three integers defined modulo 3. Of course, this ansatz follows Cardy's 
solution, but we cannot rely on his derivation because, in our case, only a 
subset of the terms $\chi_i(q)\chi_{j}(\bar q)$ 
in the partition function couples to A-type Ishibashi states, in contrast to 
the simple situation of section 3.4. That the boundary states (4.16) 
nevertheless satisfy Cardy's conditions can be checked easily: 
\eject\noindent
$$\eqalignno{
Z^A_{\alpha \tilde \alpha}(q) 
&\equiv {}_A\langle  \tilde \alpha| \; 
\tilde q^{L_0^{\rm tot} -{c\over24}} \;
|\alpha\rangle_A
&\cr
&= {1\over\sqrt{3}} 
\sum_{ { {\bf a}\atop {\scriptscriptstyle \Sigma a_i \equiv 0 ({\rm mod}\,3)}} } 
\omega^{-{\bf a}(\alpha-\tilde\alpha)} \,\chi_{{\bf a}}(\tilde q) 
= {1\over9} \sum_{{\bf c}}
\sum_{{{\bf a}\atop {\scriptscriptstyle \Sigma a_i \equiv 0 ({\rm mod}\,3)}}} 
\omega^{-{\bf a}({\bf c} + \alpha-\tilde\alpha)} \,\chi_{{\bf c}}(q) 
&\cr
&= \sum_{{\bf c}}
\delta^{(3)}_{c_1-c_3, \alpha_3-\alpha_1-\tilde\alpha_3+\tilde\alpha_1}\,
\delta^{(3)}_{c_2-c_3, \alpha_3-\alpha_2-\tilde\alpha_3+\tilde\alpha_2}\,
\chi_{{\bf c}}(q)     \ ;
&(4.17)\cr}$$
in the second line, we have used 
$$
{}_A\langle\!\langle {\bf a'}| \,\tilde q^{L_0^{\rm tot} -{c\over24}}\,
|{\bf a} \rangle\!\rangle_A = \delta_{{\bf a'},{\bf a}}\, \chi_{{\bf a}}(\tilde q)\ ,
$$
and the modular $S$-matrix (4.14); finally,  
we have carried out the two independent summations over $a_1$ and $a_2$, 
the third roots of unity yielding the periodic Kronecker symbols $\delta^{(3)}$.  
\hbn
Note that in (4.17) we did not insert the CPT operator $\Theta$ into the ``cylinder 
amplitude'', in contrast to our general discussion in section 3 and to our 
treatment of the full Gepner models below: The reason is that for simplicity 
we want to work with full representations of the $N=2$ super Virasoro algebra 
and at the same time restrict ourselves to the NS sector. Since the $N=2$ algebra 
contains fermionic generators, insertion of $\Theta$ in front 
of $|\tilde \alpha\rangle$ 
would break up the full super Virasoro representations into ones of the bosonic 
subalgebra, and the modular $S$-transformation would then force us to introduce 
the R sector as well. Since all that would only blur the salient points of the 
example, we have simply left out the CPT operator for the moment. 
\sn
Not too surprisingly, the ansatz (4.16) yields a sum of $\a^{\otimes}$-characters 
with  positive integer coefficients: $\a^{\otimes}$-symmetry in 
the open string system  $Z^A_{\alpha \tilde \alpha}(q)$ is preserved, and 
Cardy's constraints are satisfied.  But  the 
resulting boundary CFT contains unwanted states with non-integer 
U(1) charge unless we require 
$$
\alpha_1+\alpha_2+\alpha_3 =\tilde\alpha_1+\tilde\alpha_2+\tilde\alpha_3\ .
\eqno(4.18)$$
This ``compatibility condition'' between two boundary states arises essentially 
because the terms of the partition function coupling to A-type Ishibashi states 
do not close under the modular $S$-transformation. Note, however, 
that (4.18) holds for $\alpha=\tilde\alpha$. 
\sn
We have normalized the $|\alpha\rangle_A$ such that the vacuum representation 
occurs precisely once in  $Z^A_{\alpha\alpha}(q)$, cf.\ \q{28}. 
To arrive at his solution (3.32) for the boundary states of a rational CFT, 
Cardy imposed the even stronger requirement that there 
exists a special boundary state $|\tilde 0\rangle$ which has the property that 
only fields in the vacuum sector propagate in its presence, i.e.\ 
$n_{\tilde0 \tilde 0}^i = \delta_{i,0}$. 
\hbn
None of our solutions (4.16) meets this requirement. Instead, 
$$
Z^A_{\alpha\alpha}(q) = \chi_0(q)^3+\chi_1(q)^3+\chi_2(q)^3
\eqno(4.19)$$
for all  the $|\alpha\rangle_A$ in (4.16), the contributions 
corresponding to highest weight states with total dimensions and charges 
$(0,0), (1/2,1), (1/2,-1)$, respectively: $Z^A_{\alpha\alpha}(q)$ is the 
vacuum character of a free complex fermion. Furthermore, 
the spectrum in any configuration $(\alpha\,\tilde\alpha)$ of 
two compatible ``branes'' carries a representation of 
this extended symmetry algebra. 
\hbn
This is easy to understand 
since the full chiral symmetry algebra of the $(k=1)^3$ model 
in fact contains a free complex fermion $\psi(z)$, we can e.g.\ 
identify highest weight states as  $|1,1,1\rangle = \sqrt{i} 
\psi_{-{1\over2}}|{\rm vac}\rangle$, $|2,2,2\rangle = \sqrt{i} 
\psi^*_{-{1\over2}} |{\rm vac}\rangle$. With the help of the 
 $(k=1)^3$ fusion rules one can show that the boundary states (4.16)
satisfy 
$$
 \psi_r |\alpha\rangle_A =
- i\, \omega^n \,\overline{\psi}{}_{-r}^{\,*}|\alpha\rangle_A
\eqno(4.20)$$
with $n= \sum\alpha_i$. This means that the fermionic symmetry is indeed 
preserved by these boundary states. (For similar observations of extended 
symmetries in boundary CFT see \q{36}). 
\mn   
This remark allows us to make contact to the results of \q{17}, where 
the $(c,c)$ parts of the boundary states for the $(k=1)^3$ orbifold were 
given in terms of free fermion and free boson coherent states. Denote 
by $\psi, \partial X$ the left-moving complex fermion resp.\ boson with 
adjoints $\psi^*,\partial X^*$ and modes $\psi_r, \alpha_m$ etc.; 
analogously for the right-movers.  
Then the coherent A-type boundary states of Ooguri et al.\  read 
$$
|B_n \rangle = |B_n \rangle_X\, |B_n \rangle_{\psi} 
\eqno(4.21)$$ with 
$$\eqalignno{
|B_n \rangle_{\psi} &= \exp\bigl\lbrace 
i \,\omega^n \sum_{r>0} \psi_{-r}\overline{\psi}_{-r} + 
i \, \omega^{-n} \sum_{r>0} \psi^*_{-r}\overline{\psi}{}^{\,*}_{-r} 
 \bigr\rbrace\,||{\rm vac}\rangle \ , 
&(4.22)\cr
|B_n \rangle_X &= \exp\bigl\lbrace  
- \omega^n \sum_{m>0}{1\over m}\, \alpha_{-m}\overline{\alpha}_{-m} 
-  \omega^{-n} \sum_{m>0} {1\over m}\,\alpha^*_{-m}\overline{\alpha}{}^{\,*}_{-m} 
 \bigr\rbrace\,|{\rm vac}\rangle \ ; 
&(4.23)\cr
\cr}$$
they are obtained, in a slight generalization of the procedure reviewed in 
section 2,  as solutions of the Ishibashi conditions 
$$\eqalignno{
&\bigl( \psi_r + i\, \omega^n \,\overline{\psi}{}_{-r}^{\,*}
\bigr)\,|B_n \rangle_{\psi} = 0 \ , 
&(4.24)\cr
&\bigl( \alpha_m + \omega^n \overline{\alpha}{}^{\,*}_{-m}  
\bigr)\,|B_n \rangle_{X} = 0 \ ,
&(4.25)\cr}
$$
for $n=0,1,2$. The conditions for $|B_n \rangle_{X}$, which are just the 
superpartners of the ones for $|B_n \rangle_{\psi}$, have a classical geometrical 
interpretation as Dirichlet and Neumann conditions on the torus target: 
The case $n=0$ describes a 1-brane (wrapped around a supersymmetric cycle) 
extending in the $X = $ real direction, whereas in the $n=1,2$ cases 
this 1-brane (or supersymmetric cycle) is rotated by $2n\pi/3$ 
(corresponding to the $\Z_3$-symmetry of the torus), see \q{17}. 
\sn
Now let us compare our boundary states (4.16) to the coherent states (4.21) 
of Ooguri et al.: Eqs.\ (4.24) and (4.20) show that they obey the same 
Ishibashi conditions; as for the coefficients in front of the  Ishibashi states, 
we can use the fact that the Ishibashi state associated 
to an irreducible representation is unique up to normalization: We 
only need to compare the coefficients of the $N=2$ highest weight 
states $|1,1,1\rangle \otimes |1,1,1\rangle = 
i \psi_{-{1\over2}}\overline{\psi}{}_{-{1\over2}}|{\rm vac}\rangle$ 
and $|2,2,2\rangle \otimes |2,2,2\rangle = 
i \psi^*_{-{1\over2}}\overline{\psi}{}^*_{-{1\over2}}|{\rm vac}\rangle$
in (4.16) and (4.21). In both formulas, they are given 
by $\omega^n$ resp.\ $\omega^{-n}$ with $n= \sum_i \alpha_i$, 
establishing agreement. The bosonic part of the boundary state is 
of course hidden in the $N=2$ families of the fermions since 
$\partial X \sim G^-_{-{1\over2}} \psi$. 
\sn
In summary, our boundary state construction, which relied on the 
requirement that the extended algebra ${\cal A}^{\otimes}$ is 
preserved, automatically led to the coherent states which were 
used in the (geometric) $\sigma$-model approach of Ooguri et al. 
We take this as encouragement to construct analogous ``rational'' 
boundary states in arbitrary Gepner models, see section 4.3. In the 
(non-toroidal) cases beyond  $(k=1)^3$, comparisons to geometric results 
would be less direct because the $\sigma$-model interpretation is 
less explicit there. 
\hbn
We remark that open string theories assiciated to some Gepner models 
(toroidal orbifolds) were discussed in \q{48}; there, however, only 
the Chan-Paton structure guaranteeing tadpole cancellation was 
determined, which does not require explicit knowledge of the 
boundary states. 

\mn
To complete our analysis of the simplified  $(k=1)^3$ model, we compute, 
in the same way as before,  boundary states satisfying 
B-type conditions (3.19) for each component super Virasoro algebra. 
Since this requires $q^{(i)} =  - \overline{q}{}^{(i)}$ for $i=1,2,3$, 
only 3 of the 27 sectors in (4.15) provide B-type Ishibashi states, 
namely those which correspond to ``charge conjugate terms'' 
$\chi_i(q) \chi_{i^+}(\bar q)$ in the partition function, as opposed 
to the diagonal terms coupling to A-type Ishibashi states; 
cf.\ the general discussion in section 3.3.  
\hbn
The B-type boundary states can be written as 
$$
|\alpha\rangle_B = {3^{1/4}} 
\sum_{{\bf b}= (b,b,b)} 
 \omega^{-b(\alpha_1+\alpha_2+\alpha_3)}
|{\bf b} \rangle\!\rangle_B
\eqno(4.26)$$
and the same calculations as before lead to boundary CFT partition functions 
$$
Z^B_{\alpha \tilde \alpha}(q) 
\equiv {}_B\langle  \tilde \alpha| \; 
\tilde q^{L_0^{\rm tot} -{c\over24}} \;
|\alpha\rangle_B 
=\sum_{ {\bf c}= (c_1,c_2,c_3)} \delta^{(3)}_{\Sigma c_i, \Sigma \tilde \alpha_i
-\Sigma \alpha_i} \; \chi_{{\bf c}} (q) \ ;
\eqno(4.27)
$$
to obtain only states  with integer U(1) charges, the same conditions (4.18) 
as in the A-type case have to be satisfied. The spectrum described by  
$Z^B_{\alpha \tilde \alpha}(q)$ is, however, different from that in 
$Z^A_{\alpha \tilde \alpha}(q)$. 
\mn
We can also ask what the spectrum of a brane configuration 
consisting of an A-type and a B-type boundary state looks 
like. To perform the calculation, we have to realize that 
the ``inner product'' (regularized with $\tilde q$-damping 
factors as usual) of an A-type Ishibashi state and a B-type 
Ishibashi state vanishes in almost all cases, simply because 
the left-moving basis states in an Ishibashi state (3.11,16) 
are tensored to different right-moving states in the A-type 
and B-type case. More precisely, 
we find for the orbifolded $(k=1)^3$ model 
$$
{}_A\langle\!\langle {\bf a}| \,\tilde q^{L_0^{\rm tot} -{c\over24}}\,
|{\bf b} \rangle\!\rangle_B = 
\delta_{{\bf a},{\bf b}}\, \delta_{{\bf b}, (0,0,0)} \,
{\rm Tr}_{{\cal H}_{(0,0,0)}} \bigl\lb 
V_{\Omega_M} \,\tilde q^{L_0^{\rm tot} -{c\over24}} \bigr\rb
\eqno(4.28)$$
where $V_{\Omega_M}$ is the unitary operator that implements the mirror 
automorphism on the vacuum Hilbert space; on 
Poincar\'e-Birkhoff-Witt vectors,  $V_{\Omega_M}$ acts as 
$$\eqalignno{
V_{\Omega_M} \ L_{n_1} &\ldots  L_{n_i}\,  J_{m_1}\ldots  J_{m_j} \, 
     G^+_{r_1} \ldots G^+_{r_k}\,  G^-_{s_1} \ldots G^-_{s_l}\, |0,0\rangle 
&\cr
&=\  L_{n_1}\ldots  L_{n_i}\,  (-J_{m_1})\ldots  (-J_{m_j})\,  
     G^-_{r_1} \ldots G^-_{r_k}\,  G^+_{s_1} \ldots G^+_{s_l}\, |0,0\rangle \ ,
&(4.29)\cr}$$
compare eq.\ (3.18) in section 3.3. Therefore, the excitation 
spectrum of a brane configuration consisting of an A-type 
and a B-type boundary state should be computed from 
$$
Z^{AB}_{\alpha \tilde \alpha}(q) \equiv {}_A\langle  \tilde \alpha| \; 
\tilde q^{L_0^{\rm tot} -{c\over24}} \;
|\alpha\rangle_B 
= {\rm Tr}_{{\cal H}_{{\rm vac}}} \bigl\lb 
V_{\Omega_M}\,\tilde q^{L_0^{\rm tot} -{c\over24}} \bigr\rb\ ,
\eqno(4.30)$$
independently of the choice of $|\tilde\alpha\rangle_A$ and $|\alpha\rangle_B$. 
To check whether Cardy's conditions are satisfied for such a configuration, 
we would have to determine the modular transformation properties of 
the trace in (4.30). We do not indulge into this computation, because in 
the true Gepner models discussed below, the partition functions 
$Z^{AB}_{\alpha\tilde\alpha}$ vanish anyway because of space-time supersymmetry, 
enforced by the GSO projection. Nevertheless, it is an interesting fact 
that traces of automorphisms of the chiral symmetry algebra emerge 
naturally in the framework of boundary CFT D-branes. This should e.g.\ 
provide a general picture of supersymmetry breaking configurations. 

\bn\sn
{\bf 4.3 Boundary states for arbitrary Gepner models}
\mn
After this warm-up example, we discuss full-fledged Gepner models, 
taking into account R-sectors and GSO projection. We first 
have to fix the Ishibashi conditions to be imposed on the 
boundary states. As in the previous subsection, we want to use 
Cardy's results as far as possible and thus treat the internal part of 
the Gepner models as rational theories, again. This means that we 
have to choose the boundary conditions in such a way as to preserve 
an extended symmetry algebra $\cal A$, the obvious choice being 
the algebra generated by all 
the $N=2$ Virasoro algebras of the $r$ internal component theories, 
together with the SO$(d)_1$ of the external fermions. (We will 
in the following ignore the external bosons; their coherent boundary 
states, describing ordinary classical D-branes, just multiply 
the boundary states computed below). 
\sn
In the generic case, i.e.\ if the levels $k_j$ of the internal 
minimal models are pairwise different, the only way to maintain the 
tensor product symmetry in the presence of a boundary state is to require 
A-type or B-type boundary conditions as in eqs. (3.20,19) for each 
set of super Virasoro generators ${L^{(j)}_n,J^{(j)}_n,G^{\pm\,{(j)}}_r }$ 
separately. In special cases, when $k_{j_1}=k_{j_2}$, we could also glue the 
left-moving generators of subtheory $j_1$ to the right-moving 
generators of subtheory $j_2\,$, using permutation automorphisms, 
but we will not work out these boundary states here. 
\mn
We first discuss the simpler case where {\sl A-type} Ishibashi 
conditions (3.20) are imposed on each of the internal sub-Virasoro 
algebras and on the external fermions. 
A-type conditions imply that only left-right representations 
$\h_i\otimes {\h}_{\bar\imath}$ with $q_i = \bar q_{\bar\imath}$ 
(and of course $h_i = \bar h_{\bar\imath}$) contribute Ishibashi states, 
in other words, we have to restrict to the diagonal part of the Hilbert space. 
In a Gepner model  partition function (4.11), each 
left-character $\chi_i(q)$ is multiplied with $\chi_i(\bar q)$ (among others) 
on the right (where it is understood that $i$ labels irreducible 
representations of the bosonic sub-algebra of the $N=2$ 
algebras rather than full $N=2$ representations). Therefore, 
all the tensor product $A$-type Ishibashi states 
$|{\mathbf \lambda} ,{\mathbf \mu}\rangle\!\rangle_A$ (in an obvious notation) 
can occur in the boundary state provided $({\mathbf \lambda} ,{\mathbf \mu})$ 
occurs on the (left-moving) closed string spectrum. 
\sn
We make the following ansatz for ``rational'' A-type boundary states in 
Gepner models: 
$$
|\alpha\rangle_A \equiv |S_0; (L_j,M_j,S_j)_{j=1}^r \rangle_A 
= {1\over\kappa^A_{\alpha}} \sumlamube  
\;B^{{\mathbf \lambda} ,{\mathbf \mu}}_{\alpha} \,
|{\mathbf \lambda} ,{\mathbf \mu}\rangle\!\rangle_A   
\eqno(4.31)
$$
where $S_0, L_j, M_j, S_j$ are integer labels; the summation is over states 
satisfying the ``$\beta$-constraints'' as in Gepner's partition function (4.11),
$\kappa^A_{\alpha}$  is some normalization constant to be determined later, 
and the coefficients in front of the Ishibashi states 
are given by
$$
B^{{\mathbf \lambda} ,{\mathbf \mu}}_{\alpha} = 
(-1)^{{s_0^2\over2}} e^{-i\pi {d\over2} {s_0S_0\over2} } 
\prod_{j=1}^r 
{\sin \pi {(l_j+1)(L_j+1)\over k_j+2} \over
\sin^{{1\over2}} \pi {l_j+1\over k_j+2}} \,
e^{i\pi {m_jM_j\over k_j+2} } \,
e^{-i\pi {s_jS_j\over2} } \ .
\eqno(4.32)$$
Except for the sign $(-1)^{{s_0^2\over2}}$, these coefficients are 
chosen as in Cardy's solution (3.32) for the tensor product of minimal 
models and external SO$(d)_1$ factor but {\sl before} orbifolding and charge 
projections.  Moreover, after the GSO-projection we no longer deal with a 
genuine CFT anyway, thus we cannot simply rely on Cardy's general 
arguments but rather have to verify explicitly that the boundary states 
(4.31) lead to acceptable open string spectra in $Z_{\alpha \tilde\alpha}(q) = 
\langle \Theta \tilde\alpha|\, \tilde q^{L_0-{c\over24}}\, | \alpha\rangle$.  
The computations are straightforward except for the $\beta$-constraints 
in the summation. We calculate 
$$\eqalign{
 Z^A_{\alpha \tilde\alpha}(q) &=  
{1\over\kappa^A_{\alpha}\kappa^A_{\tilde\alpha}} 
\sumlamube \sumlamubetil\;
B^{{\tilde{\mathbf \lambda}} ,{\tilde{\mathbf \mu}}}_{\tilde\alpha}
B^{{\mathbf \lambda} ,{\mathbf \mu}}_{\alpha}\;
{}_A\langle\!\langle {\mathbf \tilde\lambda} ,-{\mathbf \tilde\mu}| 
\,\tilde q^{L_0-{c\over24}}\;
|{\mathbf \lambda} ,{\mathbf \mu}\rangle\!\rangle_A 
\cr
&=
{1\over\kappa^A_{\alpha}\kappa^A_{\tilde\alpha}} \sumlamube \sumlamuevpr
B^{{\mathbf \lambda} ,{\mathbf -\mu}}_{\tilde\alpha}
B^{{\mathbf \lambda} ,{\mathbf \mu}}_{\alpha}
S^{\,{\rm f}}_{s_0,s'_0} \prod_{j=1}^r 
S^{\ k_j}_{(l_j,m_j,s_j),(l'_j,m'_j,s'_j)} 
\;\chi^{{\mathbf \lambda'}}_{{\mathbf \mu'}} (q)
\cr}$$
where $\sum\!{}^{{}^{\rm ev}}$ denotes summation over the full range (4.3) 
with $l'_j+m'_j+s'_j \in2\Z$ as the only constraint, and $S^{\,{\rm f}}$
and $S^{\,k_j}$ are the modular $S$-matrices of the external fermions 
resp.\ the $j\,$th minimal model, see eqs.\ (4.12,13). 
\hbn
In order to compute the prefactor of 
$\chi^{{\mathbf \lambda'}}_{{\mathbf \mu'}}(q)$ in $Z_{\alpha \tilde\alpha}(q)$, 
we introduce Lagrange multipliers $\nu_0$, $\nu_j$, 
$j=1,\ldots,r$, for the charge constraint and the $\beta_j$-conditions 
and rewrite 
$$
\sumlamube
= \sumlamuev {1\over K}\sum_{\nu_0=0}^{K-1} e^{ i\pi \nu_0(q_{\rm tot}-1)} 
\prod_{j=1}^r \;{1\over2}\!\sum_{\nu_j=0,1} e^{i\pi \nu_j(s_0+s_j) }
$$
with $K = {\rm lcm}(4,2k_j+4)$ and the total U(1) charge  
$q_{\rm tot} = 2\beta_0\sbullet {\mathbf \mu} $
as in eq.\ (4.9). Now, the summations 
over $s_0,l_j,m_j,s_j$ are independent of each other (except for the  
$l_j+m_j+s_j$ even constraint, which is easy to handle) and can be carried 
out directly to give
$$\eqalignno{
 Z^A_{\alpha \tilde\alpha}(q) &= n^A_{\alpha\tilde\alpha}
\sumlamuevpr\ 
\sum_{\nu_0=0}^{K-1}\  
\sum_{\nu_1,\ldots,\nu_r=0,1}\  (-1)^{s_0'+S_0-\tilde S_0} \ 
\delta^{(4)}_{s_0',2+\tilde S_0-S_0 -\nu_0- 2 \Sigma \nu_j}\phantom{xxx}
&\cr
&\qquad\qquad\times\ \prod_{j=1}^r N^{l_j'}_{L_j \tilde L_j} \,
\delta^{(2k_j+4)}_{m_j',\tilde M_j-M_j -\nu_0}\,
\delta^{(4)}_{s_j',\tilde S_j-S_j -\nu_0-2\nu_j}
\;\chi^{{\mathbf \lambda'}}_{{\mathbf \mu'}} (q)
&(4.33)\cr}$$
Here, $N_{l\,l'}^{l''}$ denote the fusion rules of the SU(2)$_k$ WZW 
model, which arise from the $l_j$ summations via the Verlinde 
formula. The symbol $\delta^{(p)}_{r,s}$ means that 
$r \equiv s\ ({\rm mod}\,k)$. Finally, the normalization is 
$$
n^A_{\alpha\tilde\alpha} = {1\over\kappa^A_{\alpha}\kappa^A_{\tilde\alpha}} \,
2^{{r\over2}+1} {2(k_1+2)\cdots(k_r+2)\over K} \ .
$$
\sn
{}From eq. (4.33) we see that with the (minimal) normalization 
$$
\kappa^A_\alpha = 2 \Biggl( 2^{{r\over2}} {(k_1+2)\cdots(k_r+2)\over K}
\Biggr)^{{1\over2}}
\eqno(4.34)$$
our boundary states (4.31,32) indeed satisfy Cardy's conditions -- 
suitably modified for the supersymmetric setting: The sign in (4.33) simply 
distinguishes space-time bosons from space-time fermions. 
\hbn
But there are again additional string theory requirements 
which impose restrictions on the integers $(S_0, (L_j,M_j,S_j))$ 
in the ansatz (4.32): The spin structures of the component theories in 
$Z_{\alpha\tilde\alpha}(q)$ should be coupled as in the closed string case
(states in NS$^{\otimes(r+1)}$ or R$^{\otimes(r+1)}$ only), therefore 
we must require 
$$
S_0-\tilde S_0 \;{\buildrel {\rm !}\over \equiv}\; S_j - \tilde S_j\ ({\rm mod}\,2)
\eqno(4.35)$$ 
for all $j=1,\ldots,r$ so as to have $\beta_j\sbullet \mu' \in \Z$ 
for all states in $Z_{\alpha\tilde\alpha}(q)$. 
\hbn
For the excitation spectrum of the brane configuration $(\alpha\,\tilde\alpha)$ 
described by $Z_{\alpha\tilde\alpha}(q)$ to be supersymmetric, only states 
with odd total charge should be present. From the $\delta\,$s in 
eq.\ (4.33) one finds that 
$$
2 \beta_0\sbullet \mu' \equiv Q(\alpha-\tilde\alpha) + 1 
\;{\buildrel {\rm !}\over \equiv}\; 1 \ ({\rm mod}\,2)
\eqno(4.36{\rm a})
$$
with 
$$
Q(\alpha-\tilde\alpha) := - {d\over2}{S_0-\tilde S_0 \over2}
-\sum_{j=1}^r {S_j-\tilde S_j\over2}
+\sum_{j=1}^r {M_j-\tilde M_j\over k_j+2}\ .
\eqno(4.36{\rm b})$$
For $Q(\alpha-\tilde\alpha)$ even, only states with $q_{\rm tot}=\pm1,\pm3,\ldots$ 
contribute; since $h\geq |q|/2$ in unitary representations of the $N=2$ 
super Virasoro algebra, all states in $Z_{\alpha\tilde\alpha}(q)$ have conformal 
weight $1/2$ or higher: The spectrum is tachyon-free and stable. 
\sn
In particular, requirements (4.35,36) are satisfied for two identical 
branes $\alpha=\tilde\alpha$ given by A-type boundary states (4.31,32), 
meaning that the excitation spectrum of a single such brane is 
supersymmetric and stable. 
\hbn
On the other hand, (4.35,36) imply that not any two boundary states are 
``compatible'' with each other. Rather, we obtain groups of mutually 
compatible boundary states; configurations made up of two branes from different 
groups lead to spectra violating supersymmetry (and stability). E.g., a 
``brane-antibrane system'' $(\alpha\,\alpha^+)$ with  $|\alpha^+\rangle := 
\Theta\,|\alpha\rangle$ has a tachyon in its spectrum. 

\mn
\eject\noindent
Since half of the space-time supersymmetry is preserved by our boundary 
states, we may conclude that they describe BPS-states. A further check 
of this property would be to show that $Z_{\alpha \alpha}(q)$ in fact
vanishes. 
\hbn
Since the $\chi^{{\mathbf \lambda'}}_{{\mathbf \mu'}} (q)$ consist of 
products of theta functions, identities for the latter should in principle 
allow to verify $Z_{\alpha \alpha}(q)=0$ in general (analogous to 
Polchinski's computation for the flat case \q{4,5}, which rests on 
Jacobi's ``abstruse identity''). As yet,  we have 
merely expanded $Z_{\alpha \alpha}(q)$ for some models up to some 
power in $q$, finding indeed zero. 

\bn
According to the results of \q{17}, the above A-type boundary states
should, via the correspondence between Gepner models and Calabi-Yau 
manifolds mentioned before, describe D-branes wrapping around 
 middle-dimensional supersymmetric cycles of the CY manifold. Moreover, using 
topological twisting of the $N=2$ theory, Ooguri et al.\ have argued 
that the coefficients $B^{{\mathbf \lambda} ,{\mathbf \mu}_c}_{\alpha}$ 
in front of Ishibashi states $|{\mathbf \lambda} ,{\mathbf \mu}_c\rangle\!\rangle_A$ 
corresponding to chiral primaries (fields with $q=2h$) 
are independent of the K\"ahler moduli 
and can be, in the large volume limit, expressed in geometric terms: 
$$
B^{{\mathbf \lambda} ,{\mathbf \mu}_c}_{\alpha} 
= \int_{\gamma_{\alpha}} \omega_{{\mathbf \lambda} ,{\mathbf \mu}_c}
\eqno(4.37)$$
where $ \omega_{{\mathbf \lambda} ,{\mathbf \mu}_c}$ is a differential 
form associated to the chiral primary $\phi_{{\mathbf \lambda} ,{\mathbf \mu}_c}$
via topological twisting, see \q{17} for the details. 
In a Gepner model, the chiral primaries correspond 
to irreducible representations labeled by 
$$
({\mathbf \lambda},\,{\mathbf \mu}_c)  =
 ({\mathbf \lambda},\,(0;{\mathbf \lambda};0,\ldots,0))
$$ 
and the corresponding coefficients in the boundary state $|\alpha\rangle_A$ read 
$$
B^{{\mathbf \lambda} ,{\mathbf \mu}_c}_{\alpha} =
{1\over\kappa^A_\alpha}\; \prod_{j=1}^r 
{\sin \pi {(l_j+1)(L_j+1)\over k_j+2} \over
\sin^{{1\over2}} \pi {l_j+1\over k_j+2}} \,
e^{i\pi {l_jM_j\over k_j+2} } \ .
\eqno(4.38)$$
But beyond this ``geometric part'' of the boundary states, our formula 
(4.32) contains the non-chiral contributions, too. 
\sn
As was recalled in section 2, eqs.\ (4.31,32) moreover allow 
to determine the tension and RR charges of the branes described by 
these boundary states. One merely has to project $|\alpha\rangle_A$ 
onto the massless closed string state in question (and take into account 
universal prefactors from string amplitudes). While we could arrive 
at a closed formula for (rational) boundary states for Gepner model, 
the massless closed string spectrum does depend sensitively on the concrete 
model, and we will not indulge into further case studies in this article. 
\hbn
Similar remarks apply to the excitation spectrum of brane configurations, 
which can be extracted from $Z_{\alpha \alpha}(q)$. 
Since its massless part determines the field content of the low-energy
effective field theory associated to the configuration, explicit 
knowledge might be interesting for the study of gauge theories in four   
dimensions. 

\bn
Let us now discuss {\sl B-type} boundary states with B-type Ishibashi 
conditions (3.19) imposed on each sub-theory. The U(1) charges of the left- 
and right-moving highest weight states must satisfy $q_i=-\bar q_{\bar\imath}$
(and $h_i = \bar h_{\bar\imath}$), and the B-type Ishibashi states
couple to ``charge conjugate'' parts  $\h_i\otimes {\h}_{i^+}$ of the 
bulk Hilbert space, as discussed in section 3.3. A little bit of 
calculation shows that a term $\chi^{{\mathbf \lambda}}_{{\mathbf \mu}} (q)
\chi^{{\mathbf \lambda}}_{{\mathbf -\mu}} (\bar q)$ occurs in 
the Gepner partition function (4.11) 
precisely for those ${\mathbf \mu}$ which satisfy 
$$
m_j \equiv b\ ({\rm mod}\, k_j+2) 
\eqno(4.39)$$
for some $b=0,1,\ldots,{K\over2}-1$ and for all $j$. It is those Ishibashi 
states that contribute to the sum in the ansatz for the B-type boundary states 
$$
|\alpha\rangle_B\equiv |S_0; (L_j,M_j,S_j)_{j=1}^r \rangle_B 
= {1\over\kappa^B_{\alpha}} \sumlamubebe  
\;B^{{\mathbf \lambda} ,{\mathbf \mu}}_{\alpha} \,
|{\mathbf \lambda} ,{\mathbf \mu}\rangle\!\rangle_B   
\eqno(4.40)
$$
with  coefficients $B^{{\mathbf \lambda} ,{\mathbf \mu}}_{\alpha}$ as 
before (4.32). Note that, generically, Gepner models possess less 
B-type than A-type Ishibashi states; as a consequence, the excitation 
spectra of B-type brane configurations will typically be richer than 
those of A-type branes. 
\sn
One can now perform the same calculation as before in order to test 
this ansatz; because of the restricted $m_j$-range (4.39), there  
are slight differences compared to A-type boundary conditions, but 
again Lagrange multipliers can be used to disentangle all summations. 
For the partition function describing the excitation spectrum of 
a configuration of two B-type branes as in eq.\ (4.40) one obtains 
$$\eqalignno{
 Z^B_{\alpha \tilde\alpha}(q) &= n^B_{\alpha\tilde\alpha}
\sumlamuevpr\ 
\sum_{\nu_0=0}^{K-1}\  
\sum_{\nu_1,\ldots,\nu_r=0,1}\  (-1)^{s_0'+S_0-\tilde S_0} \ 
\delta^{(4)}_{s_0',2+\tilde S_0-S_0 -\nu_0- 2 \Sigma \nu_j}\phantom{xx}
&\cr
&\ \times\ \delta^{(K')}_{\Sigma_{m'},0} 
\;\prod_{j=1}^r N^{l_j'}_{L_j \tilde L_j} \,
\delta^{(2)}_{m_j'+M_j-\tilde M_j +\nu_0,0}\,
\delta^{(4)}_{s_j',\tilde S_j-S_j -\nu_0-2\nu_j}
\ \chi^{{\mathbf \lambda'}}_{{\mathbf \mu'}} (q)
&(4.41{\rm a})\cr}$$
with $K' := {\rm lcm}(2k_j+4)$, $\xi_j := K'/(2k_j+4)$ and 
$$
\Sigma_{m'} := \sum_{j=1}^r \xi_j\,(m_j'+M_j-\tilde M_j + \nu_0)\ ;
\eqno(4.41{\rm b})$$
the overall factor is 
$n^B_{\alpha\tilde\alpha}= 2^{r\over2}/(\kappa^B_{\alpha}\kappa^B_{\tilde\alpha})$ 
so that we choose the normalization in (4.40) to be 
$$
\kappa^B_{\alpha}= 2^{r\over4}\ .
\eqno(4.42)$$
As in the A-type case, Cardy's conditions are always satisfied for 
our boundary states (4.40), and the remaining string requirements 
for the spectrum of a brane configuration $(\alpha\,\tilde\alpha)$ 
lead to the same conditions 
$$
S_0-\tilde S_0 \;{\buildrel {\rm !}\over \equiv}\; 
S_j - \tilde S_j\ ({\rm mod}\,2)\ , \quad
Q(\alpha-\tilde\alpha) 
\;{\buildrel {\rm !}\over \equiv}\; 0 \ ({\rm mod}\,2)
$$
on the integer labels occurring in the ansatz. 
\mn
As in the simplified $(k=1)^3$ example treated in section 4.2, we can 
take a look at systems consisting of one A-type and one B-type 
brane. For the Gepner models, we find that A-type and B-type 
boundary states do not ``exert static forces on each other'', or 
$$
Z^{AB}_{\alpha \tilde \alpha}(q) = 0\ ,
$$ 
simply because of the GSO projection: Again, only representations with total 
charge $q=0$ could contribute to the trace occurring as in (4.28), but 
the GSO projection leaves only states with odd integer charge in 
the spectrum. 

\bn\bn
{\bf 5. Summary and outlook} 
\mn
We have seen that the language of boundary CFT 
allows for a general conceptual formulation of (static) D-branes, 
i.e.\ of non-perturbative sectors of a closed string theory. Strong 
constraints make it possible to arrive at concrete and complete 
formulas even in the relatively complicated case of Gepner models. 
In contrast, the geometric methods seem to yield only part of 
the boundary states, thus part of the excitation spectrum 
of brane configurations remains undetermined.
In the CFT approach, we 
have explicit control of the excitation spectrum: Its symmetries are 
manifest, and so is stability. As a consequence, we discover classes of 
mutually compatible branes. Given the boundary state, it is 
straightforward to compute the coupling constants (tensions, RR charges) 
of the low-energy effective field theory, simply by sandwiching the boundary 
state with the relevant closed string states. 
\sn
An obvious question that has been left open here is whether the additional 
sewing constraints of boundary CFT \q{30,35,36} eliminate some of the 
Gepner model boundary states we have found. There are indications 
that this is not the case, but one should certainly clarify this point;  
the general methods developed in \q{37}, see also \q{49}, should 
prove efficient in this context. 
\sn
Gepner models can also be formulated starting from modular invariant 
partition functions that are non-diagonal in their SU(2) part. Using 
the methods of \q{36,37}, our construction can be extended to 
this case, too. Similarly, one could discuss other string compactifications, 
in particular the Kazama-Suzuki models whose Ishibashi conditions were 
analyzed in \q{50}.  
\sn
Beyond that, one can try to make further use of the fact that the 
boundary state approach gives full control of the excitation 
spectrum of brane configurations: In particular, it would be 
interesting to search for configurations that break part of 
the supersymmetry; in the flat case, this can e.g.\ be achieved 
by letting the branes intersect at certain angles, 
more generally by putting boundary states with different 
Ishibashi conditions on both ends of an open string. 
In general, this gives contributions to $Z_{\alpha\tilde\alpha}(q)$ 
like the ${\rm Tr}_{{\cal H}_{\rm inv}} 
V_{\Omega} \tilde q^{L_0 - {c\over 24}}$ in section 4.2. 
\sn
To study such problems, and for general reasons, it would be 
advantageous to have moduli in the boundary states, like the location 
of the classical brane in the flat target case. Even more interesting 
results arise from varying the moduli, e.g.\ from computing string 
amplitudes for D-branes in relative motion \q{51}. 
It seems that all this requires to consider non-rational situations, 
which in general is much harder from the CFT point of view. We hope 
that one can make progress e.g.\ by deforming rational boundary 
states like the ones we have constructed above away from the 
rational point. 
\sn
Another immediate question is how one can describe, in CFT terms, 
configurations with three or more branes. It should be possible to 
extend the formulation in \q{20} from the flat target case 
to our generalized D-branes. Furthermore, it appears that the 
boundary state formalism also allows to investigate 
bound states of (strings and) branes, cf.\ \q{52} and the recent work 
\q{53}; hopefully, one can use CFT methods to prove existence of 
the marginal bound states important for M-theory.
\sn
Definitely, one should try to establish closer contact to the geometric 
approach to D-branes. While the latter has certain short-comings 
as far as completeness of the D-brane description or calculation of concrete 
spectra is concerned, it is of course much more effective in controlling 
moduli. 
\sn
We believe, however, that a true ``geometric understanding'' of D-branes 
should be sought within non-commutative geometry rather than classical 
geometry. The main reason was recalled already in the introduction: String 
theory by its very design reaches beyond classical notions of space-time,   
and non-commutative geometry \q{54} is the most general and best 
developed framework to discuss ``quantized space-time''. Moreover, it 
is precisely through D-branes and their role in the matrix model 
proposal \q{55} that aspects of non-commutative geometry have appeared 
naturally in string theory. In the recent work \q{56}, this interplay 
was exploited to study matrix theory compactifications on non-commutative 
tori. Indications that D-branes are related to cohomology classes of some 
differential calculus intrinsic to string theory were given in \q{57}. 
It is certainly worthwhile to pursue these matters further. 
 
\bn\bn
{\bf Acknowledgments} 
\mn
We would like to thank M.\ Bauer, U.\ Danielsson, R.\ Dijkgraaf, 
M.\ Douglas, J.\ Fr\"ohlich, M.\ Gaberdiel, K.\ \gaw, O.\ Grandjean, 
W.\ Lerche, W.\ Nahm, R.\ Schimmrigk, C.\ Schweigert, Y.\ Stanev,  
M.\ Terhoeven and J.-B.\ Zuber 
for very useful and stimulating discussions. 
\hbn
Part of this work was done at ETH Z\"urich; we are grateful to 
the $M \cup \Phi$ group for the hospitality extended to us. 
\bn\bn\bn
\def\q#1{\cr$\lb{\rm #1}\rb$}\vfill\eject
\parindent=35pt \vsize=23.3truecm
\noindent
\halign{#\hfil&\vtop{\parindent=0pt \hsize=37.3em #\strut}\cr
\noalign{\leftline{{\bf References }}} \noalign{\vskip.4cm}
$\lb{\rm 1}\rb$ &M.J.\ Duff, R.R.\ Khuri, J.X.\ Lu, {\sl String solitons}, 
Phys.\ Rep.\ {\bf259} (1995) 213, hep-th/9412184. 
\q{2} &J.\ Dai, J.\ Polchinski, R.G.\ Leigh, {\sl New connections 
 between string theories}, Mod.\ Phys.\ Lett.\ {\bf A4} (1989) 2073.
\q{3} &J.\ Polchinski, {\sl Combinatorics of boundaries in string 
 theory}, Phys.\ Rev.\ {\bf D 50} (1994) 6041, hep-th/9407031.
\q{4} &J.\ Polchinski, {\sl Dirichlet branes and Ramond-Ramond 
 charges}, Phys.\ Rev.\ Lett.\ {\bf 75} (1995) 4724, hep-th/9510017.
\q{5} &J.\ Polchinski, {\sl TASI lectures on D-branes}, 
 hep-th/9611050.
\q{6} &E.\ Witten, {\sl String theory dynamics in various dimensions}, 
  Nucl.\ Phys.\  {\bf B443} (1995) 85, hep-th/9503124.
\q{7} &P.K.\ Townsend, {\sl The eleven-dimensional supermembrane revisited}, 
   Phys.\ Lett.\ {\bf B350} (1995) 184, hep-th/9501068.
\q{8} &E.\ Witten, {\sl Bound states of strings and D-branes}, 
 Nucl.\ Phys.\ {\bf B460} (1996) 335, hep-th/9510135.
\q{9} &S.\ Kachru, A.\ Klemm, W.\ Lerche, P.\ Mayr, C.\ Vafa,  {\sl 
   Nonperturbative results on the point particle limit of \hbox{$N=2$} heterotic 
  string compactifications}, Nucl.\ Phys.\ {\bf B459} (1996) 537, hep-th/9508155.
\q{10} &A.\ Klemm, W.\ Lerche, P.\ Mayr, C.\ Vafa, N.\ Warner, 
    {\sl Self-dual strings and $N=2$ supersymmetric field theory}, 
    Nucl.\ Phys.\ {\bf B477} (1996) 746, hep-th/9604034.
\q{11} &M.R.\ Douglas, M.\ Li, {\sl D-brane realization of $N=2$ super Yang 
      Mills theory in four dimensions}, hep-th/9604041.
\q{12} &A.\ Hanany, E.\ Witten, {\sl Type IIB superstrings, BPS monopoles, and 
    three-dimensional gauge dynamics}, Nucl.\ Phys.\ {\bf B492} (1997) 152,
  hep-th/9611230;\cr 
   &E.\ Witten, {\sl Solutions of four-dimensional field theories 
  via M theory}, Nucl.\ Phys.\  {\bf B500} (1997) 3, hep-th/9703166. 
\q{13} &W.\ Lerche, {\sl Recent developments in string theory}, hep-th/9710246.
\q{14} &M.R.\ Douglas, D.\ Kabat, P.\ Pouliot, S.H.\ Shenker, 
  {\sl D-branes and short distances in string theory}, 
  Nucl.\ Phys.\ {\bf B485} (1997) 85, hep-th/9608024.  
\q{15} &K.\ Becker, M.\ Becker, A.\ Strominger, {\sl Fivebranes, membranes and 
      non-perturbative string theory}, Nucl.\ Phys.\ {\bf B456} (1995) 130, 
      hep-th/9507158. 
\q{16} &M.\ Bershadsky, V.\ Sadov, C.\ Vafa, {\sl D-Branes and topological 
   field theories}, Nucl.\ Phys.\ {\bf B463} (1996) 420, hep-th/9511222.
\q{17} &H.\ Ooguri, Y.\ Oz, Z.\ Yin, {\sl D-branes on
 Calabi-Yau spaces and their mirrors}, Nucl.\ Phys.\ {\bf B477}
 (1996) 407, hep-th/9606112.
\q{18} &P.\ Di Vecchia, L.\ Magnea, R.\ Marotta, A.\ Lerda, R.\ Russo, 
  {\sl The field theory limit of multiloop string amplitudes}, hep-th/9611023.
\q{19} &M.\ Billo, D.\ Cangemi, P.\ Di Vecchia, {\sl
 Boundary states for moving D-branes}, Phys.\ Lett.\ {\bf B400}
 (1997) 63, hep-th/9701190.
\q{20} &M.\ Frau, I.\ Pesando, S.\ Sciuto, A.\ Lerda,
 R.\ Russo, {\sl Scattering of closed strings from many
 D-branes}, Phys.\ Lett.\ {\bf B400} (1997) 52, hep-th/9702037.
\q{21}&P.\ Di Vecchia,  M.\ Frau, I.\ Pesando, S.\ Sciuto,
 A.\ Lerda, R.\ Russo, {\sl Classical p-branes from boundary states},
 Nucl.\ Phys.\ {\bf B507} (1997) 259, hep-th/9707068.\cr
$\lb{\rm 22}\rb$&F.\ Hussain, R.\ Iengo, C.\ Nu\~nez, C.A.\ Scrucca, 
    {\sl Interaction of moving D-branes on orbifolds}, Phys.\ Lett.\ 
    {\bf B409} (1997) 101, hep-th/9706186;\quad 
   {\sl Closed string radiation from moving D-branes}, hep-th/9710049;\quad 
    {\sl Interaction of D-branes on orbifolds and massless particle emission}, 
    hep-th/9711021. 
\q{23} &C.G.\ Callan, C.\ Lovelace, C.R.\ Nappi, S.A.\ Yost, 
 {\sl Adding holes and crosscaps to the superstring}, Nucl.\ Phys.\ 
 {\bf B293} (1987) 83; \quad {\sl Loop corrections to superstring equations 
 of motion},  Nucl.Phys.\ {\bf B308} (1988) 221.
\q{24} &J.\ Polchinski, Y.\ Cai, {\sl Consistency of open superstring 
 theories}, Nucl.\ Phys.\ {\bf B296} (1988) 91. 
\q{25} &O.\ Bergman, M.R.\ Gaberdiel, {\sl A non-supersymmetric
 open string theory and S-duality}, Nucl.\ Phys {\bf B499} (1997) 183,
 hep-th/9701137; \cr
 &O.\ Bergman, M.R.\ Gaberdiel, G.\ Lifschytz, {\sl
  Branes, orientifolds and creation of elementary strings}, Nucl.\ Phys.\ 
  {\bf B509} (1998) 194, hep-th/9705130.
\q{26} &M.\ Oshikawa, I.\ Affleck, {\sl Boundary
 conformal field theory approach to the critical two-dimensional Ising model 
  with a defect line},  Nucl.\ Phys {\bf B495} (1997) 533, cond-mat/9612187.
\q{27} &J.L.\ Cardy, {\sl Conformal invariance and surface
 critical behavior}, Nucl.\ Phys.\ {\bf B240} (1984) 514; 
  \quad {\sl Effect of boundary conditions on the
 operator content of two-dimensional conformally invariant theories},
 Nucl.\ Phys.\ {\bf B275} (1986) 200.\cr
$\lb{\rm 28}\rb$&J.L.\ Cardy, {\sl Boundary conditions, fusion rules
 and the Verlinde formula}, Nucl.\ Phys.\ {\bf B324} (1989) 581. 
\q{29}&A.\ Sagnotti, {\sl Open strings and their symmetry groups}, in:\   
  Non-Perturbative Methods in Field Theory, eds.\ G.\ Mack et al., 
  Lecture Notes Carg\`ese 1987;   \cr
 &M.\ Bianchi, A.\ Sagnotti,  {\sl On the systematics of open string
  theories},  Phys.\ Lett.\ {\bf B247} (1990) 517; \cr
 &A.\ Sagnotti, {\sl Surprises in open-string perturbation theory}, 
   Nucl.\ Phys.\ Proc.\ Suppl.\ {\bf 56B} (1997) 332, hep-th/9702093.
\q{30} &J.L.\ Cardy, D.C.\ Lewellen, {\sl Bulk and boundary operators
 in conformal field theory}, Phys.\ Lett.\ {\bf B259} (1991) 274.
\q{31} &N.\ Ishibashi, {\sl The boundary and crosscap states
 in conformal field theories}, Mod.\ Phys.\ Lett.\ {\bf A4} (1989) 251; \cr
 &N.\ Ishibashi, T.\ Onogi, {\sl Conformal field theories on surfaces
 with boundaries and crosscaps}, Mod.\ Phys.\ Lett.\ {\bf A4} (1989) 161.
\q{32} &M.\ Kato, T.\ Okada, {\sl D-branes on group manifolds}, 
  Nucl.\ Phys.\ {\bf B499} (1997) 583, hep-th/9612148.  
\q{33} &W.\ Lerche, C.\ Vafa, N.P.\ Warner, {\sl Chiral rings 
   in $N=2$ superconformal theories}, Nucl.\ Phys.\ {\bf B324} (1989) 427.
\q{34} &E.\ Wong, I.\ Affleck, {\sl Tunneling in quantum wires: a boundary 
  conformal field theory approach},  Nucl.\ Phys.\ {\bf B417} (1994) 403.
\q{35} &D.C.\ Lewellen, {\sl Sewing constraints for conformal field
 theories on surfaces with boundaries}, Nucl.\ Phys.\ {\bf B372} (1992) 654.
\q{36} &G.\ Pradisi, A.\ Sagnotti, Y.S.\ Stanev, {\sl Completeness 
   conditions for boundary operators in 2d conformal field theory}, 
   Phys.\ Lett.\ {\bf B381} (1996) 97, hep-th/9603097;\cr
&{\sl Planar duality in $SU(2)$ WZW 
   models}, Phys.\ Lett.\ {\bf B354} (1995) 279, hep-th/9503207;\cr
&{\sl The open descendants of 
   non-diagonal SU(2) WZW models}, Phys.\ Lett.\ {\bf B356} (1995) 230, 
  hep-th/9506014. 
\q{37} &J.\ Fuchs, C.\ Schweigert, {\sl A classifying algebra for boundary 
  conditions}, Phys.\ Lett.\ {\bf B414} (1997) 251, hep-th/9708141.
\q{38} &D.\ Gepner, {\sl Space-time supersymmetry in compactified
 string theory and superconformal models}, Nucl.\ Phys.\ {\bf B296}
 (1987) 757.
\q{39} &B.R.\ Greene, {\sl String theory on Calabi-Yau manifolds},
          TASI lectures, hep-th/9702155.\cr
$\lb{\rm 40}\rb$&J.\ Fuchs, A.\ Klemm, C.\ Scheich, M.G.\ Schmidt,
 {\sl Gepner models with arbitrary affine invariants and the
 associated Calabi-Yau spaces}, Phys.\ Lett.\ {\bf B232} (1989) 317;\cr
 &{\sl Spectra and symmetries of Gepner models compared to
 Calabi-Yau compactifications}, Ann.\ of Phys.\ {\bf 204} (1990) 1.   
\q{41} &O.\ Grandjean, private notes. 
\q{42} &W.\ Lerche, B.\ Schellekens, N.\ Warner, {\sl Lattices and 
   strings},  Phys.\ Rep.\ {\bf 177} (1989) 1.\cr
$\lb{\rm43}\rb$&C.A.\ L\"utken, G.G.\ Ross, {\sl Taxonomy of heterotic 
  superconformal field theories},  Phys.\ Lett.\ {\bf B213} (1988) 152;\cr
  &M.\ Lynker, R.\ Schimmrigk, {\sl On the spectrum of (2,2) 
  compactification of the heterotic string on conformal field theories}, 
  Phys.\ Lett.\ {\bf B215} (1988) 681.
\q{44} &P.\ Candelas, X.C.\ de la Ossa, P.S.\ Green, L.\ Parkes, 
  {\sl An exactly 
  soluble superconformal theory from a mirror pair of Calabi-Yau manifolds}, 
  Phys.\ Lett.\ {\bf B258} (1991) 118.
\q{45} &B.R.\ Greene, M.R.\ Plesser, {\sl Duality in Calabi-Yau
 moduli spaces}, Nucl.\ Phys.\ {\bf B338} (1990) 14.
\q{46} &S.T.\ Yau (ed.),  Essays on Mirror Manifolds,  
      International Press 1992.
\q{47} &R.\ Dijkgraaf, C.\ Vafa, E.\ Verlinde, H.\ Verlinde,
 {\sl Operator algebra of orbifold models}, Commun.\ Math.
 Phys.\ {\bf 123} (1989) 485.
\q{48}&M.\ Bianchi, A.\ Sagnotti,  {\sl Twist symmetry and open string 
  Wilson lines}, Nucl.\ Phys.\ {\bf B361} (1991) 519; \cr
  &M.\ Bianchi, G.\ Pradisi, A.\ Sagnotti,  {\sl Toroidal compactification 
  and symmetry breaking in open string theories}, Nucl.\ Phys.\ 
  {\bf B376} (1992) 365; \cr
  &C.\ Angelantonj, M.\ Bianchi, G.\ Pradisi, A.\ Sagnotti, Y.S.\ Stanev, 
   {\sl Comments on Gepner models and type I vacua in string theory},  
   Phys.\ Lett.\ {\bf B387} (1996) 743, hep-th/9607229. 
\q{49} &J.\ Fuchs, C.\ Schweigert, {\sl Branes:\ from free fields to general 
  conformal field theories}, hep-th/9712257.
\q{50} &S.\ Stanciu, {\sl D-branes in Kazama-Suzuki models}, hep-th/9708166.\cr
$\lb{\rm 51}\rb$&C.\ Bachas, {\sl D-brane dynamics}, Phys.\ Lett.\ {\bf B374} 
     (1996) 37, hep-th/9511043.\cr
$\lb{\rm 52}\rb$&C.G.\ Callan, I.R.\ Klebanov, {\sl D-Brane boundary state 
  dynamics}, Nucl.\ Phys.\  {\bf B465} (1996) 473, hep-th/9511173.
\q{53} &S.\ Gukov, I.R.\ Klebanov, A.M.\ Polyakov, 
   {\sl Dynamics of $(n,1)$ Strings}, hep-th/9711112.\cr
$\lb{\rm 54}\rb$&A.\ Connes, Noncommutative Geometry, Academic Press 1994.
\q{55} & T.\ Banks, W.\ Fischler, S.H.\ Shenker, L.\ Susskind, {\sl M 
  Theory as a matrix model: a conjecture}, Phys.\ Rev.\ {\bf D55} (1997) 5112, 
  hep-th/9610043. 
\q{56} &A.\ Connes, M.R.\ Douglas, A.\ Schwarz, {\sl Noncommutative geometry 
   and matrix theory: compactification on tori}, hep-th/9711162.
\q{57} &J.\ Fr\"ohlich, O.\ Grandjean, A.\ Recknagel, 
     {\sl Supersymmetric quantum theory, non-com\-mutative 
     geometry, and gravitation},
     Les Houches Lecture Notes 1995, hep-th/9706132. 
\cr}
\bye